\documentclass[12pt,preprint]{emulateapj}

\slugcomment{AJ}

\shorttitle{High density molecular gas of NGC 1614 with ALMA}
\shortauthors{Imanishi et al.}

\begin{document}


\title{High-density Molecular Gas Properties of the 
Starburst Galaxy NGC 1614 Revealed with ALMA}   


\author{Masatoshi Imanishi\altaffilmark{1,2}}
\affil{Subaru Telescope, 650 North A'ohoku Place, Hilo, Hawaii, 96720,
U.S.A.} 
\email{masa.imanishi@nao.ac.jp}

\and

\author{Kouichiro Nakanishi\altaffilmark{1,2}}
\affil{Joint ALMA Observatory, Alonso de Cordova 3107, Vitacura 
763-0355, Santiago de Chile}

\altaffiltext{1}{National Astronomical Observatory of Japan, 2-21-1
Osawa, Mitaka, Tokyo 181-8588}
\altaffiltext{2}{Department of Astronomy, School of Science, Graduate
University for Advanced Studies (SOKENDAI), Mitaka, Tokyo 181-8588}

\begin{abstract}
We present the results of HCN/HCO$^{+}$/HNC J = 4--3 transition line  
observations of the nearby starburst galaxy NGC 1614, obtained with ALMA 
Cycle 0. 
We find that high density molecular gas, traced with these lines, shows
a velocity structure such that the northern (southern) side of the
nucleus is redshifted (blueshifted) with respect to the nuclear velocity
of this galaxy.  
The redshifted and blueshifted emission peaks are offset by
$\sim$0.6$''$ at the northern and southern sides of the nucleus,
respectively.
At these offset positions, observations at infrared $>$3 $\mu$m 
indicate the presence of active dusty starbursts,
supporting the picture that high-density molecular gas is the site of
active starbursts.  
The enclosed dynamical mass within the central $\sim$2$''$ in radius,
derived from the dynamics of the high-density molecular gas, is
$\sim$10$^{9}$ M$_{\odot}$, which is similar to previous estimates.  
Finally, the HCN emission is weaker than HCO$^{+}$ but
stronger than HNC for J = 4--3 for all starburst regions of NGC 1614, as
seen for J = 1--0 transition lines in starburst-dominated galaxies. 
\end{abstract}


\section{Introduction}

Luminous infrared galaxies (LIRGs) show strong infrared emission, 
with infrared (8--1000 $\mu$m) luminosities of L$_{\rm IR}$ $>$
10$^{11}$L$_{\odot}$, created by energy sources hidden behind dust.  
They are mostly found in gas-rich galaxy mergers \citep{sam96}.
Molecular gas in galaxy mergers is largely influenced by 
merger-induced physical processes, and obtaining observational
constraints on the spatial distribution and dynamics of molecular gas in
merging LIRGs is important to our understanding of gas-rich galaxy
merger processes.  
(Sub)millimeter interferometric observations of rotational J-transition
lines of molecular gas are a powerful tool for this purpose.  
High-sensitivity, high-spatial-resolution ($<$a few arcsec) 
interferometric observations of merging LIRGs using bright CO molecular
lines have been widely performed \citep{dow98,bry99,tru01,eva02,dow07}. 
These CO observations at low-J transitions (J = 1--0 and 2--1) have
effectively traced the low-density (10$^{1-3}$ cm$^{-3}$) molecular gas  
properties in great detail due to the low dipole moment of CO 
($\mu$ $\sim$ 0.1 debye).  
However, in merging LIRGs, the fraction of high density 
($>$10$^{4}$ cm$^{-3}$) molecular gas is much higher than in normal
quiescent star-forming galaxies \citep{sol92,gao04},  
and it is within such high-density gas that stars are actually born. 
Thus, it is vital to obtain observational constraints of the properties
of high-density molecular gas in merging LIRGs if we are to understand
the essential physical processes in gas-rich galaxy mergers.  

Observations of molecular gas with high dipole moments, such as 
HCN, HCO$^{+}$, and HNC ($\mu$ $>$ 3 debye), can effectively probe
high-density molecular gas.  
However, these molecular lines are generally much fainter than the bright
CO lines, and the spatial information of dense gas is still limited to
nearby, very bright merging LIRGs only 
\citep{aal97,cas99,ima04,nak05,ima06b,in06,ima07,ion07,wil08,ima09,aal09,
sak09,sak13}.  
From these previously performed observations, it has been shown that the
spatial distribution of high-density molecular gas is significantly
different from that of low-density molecular gas, in that high-density
gas is more concentrated in the nuclear regions of galaxies
\citep{ion04,nak05,ima07,wil08,sak13}.
In merging LIRGs, it is in the nuclear regions where very violent
processes, including starbursts (= active star-formation) and active
mass accretion onto a supermassive black hole (SMBH) occur.   
Additionally, feedback to the surrounding interstellar medium and entire
galaxy, if present, originates from these regions
\citep{hop05,spr05,dim05,hop06}.   
Spatially resolved interferometric observations using dense gas tracers 
are of particular importance in investigating the interesting nuclear
regions of merging LIRGs.   

Starburst and active galactic nucleus (AGN) activity powered by a
mass-accreting SMBH can have different effects/feedback on the
surrounding dense molecular gas at the merging LIRG's nuclei. 
It is proposed that starburst and AGN activity could be distinguishable
based on the line flux ratios of dense molecular gas tracers
\citep{koh05,ima04,ima06b,ima07,ima09,kri08} because 
(1) an AGN has an energy source with a much higher emission surface
brightness than a starburst, and thus it can heat the surrounding dust and gas
to higher temperature; and (2) an AGN emits stronger X-rays than
a starburst does. Both of these factors could alter the chemical
compositions of molecular gas in AGNs compared with starbursts 
\citep{mei05,lin06,har10}; therefore, AGNs and starbursts could exhibit
different molecular line flux ratios.  
High-spatial-resolution interferometric observations of multiple
dense gas tracers can thus be used to scrutinize the processes deep
inside the obscuring dust and gas in merging LIRG nuclei.
With the advent of ALMA, such observations are now feasible. 

NGC 1614 (z = 0.016; L$_{\rm IR}$ = 10$^{11.6}$L$_{\odot}$) is a
well-studied, nearby LIRG.
Merging signatures are clearly seen in the optical and near-infrared
(1--2.5 $\mu$m) images as long, prominent tails around a single nucleus
\citep{alo01,rot04,haa11}.  
It is classified as a starburst galaxy through optical spectroscopy
\citep{vei95,kew01,yua10}. 
The infrared 2.5--30 $\mu$m spectrum of NGC 1614 is typical of a
starburst-dominated galaxy with no AGN signature
\citep{bra06,ber09,ima10b,vai12}.   
The luminosities of the starburst-generated 3.3 $\mu$m polycyclic
aromatic hydrocarbon (PAH) emission feature and the Br$\alpha$ (4.05
$\mu$m) hydrogen emission, measured through slitless spectroscopy and
calculated relative to the total infrared (8--1000 $\mu$m) luminosity,
are both as high as in starburst-dominated galaxies \citep{ima10b},
suggesting that the observed 
luminosity of NGC 1614 is totally accounted for by the detected
starbursts, with no need for a significant AGN contribution. 
The infrared $K$-band (2.2 $\mu$m) spectrum shows a strong
stellar-origin 2.3 $\mu$m CO absorption feature \citep{rid94,alo01} and
so supports the starburst-dominated scenario.  
High-spatial-resolution infrared 8--20 $\mu$m imaging observations 
reveal spatially extended, but compact ($\sim$2$''$), starburst-heated 
dust continuum emission \citep{mil96,soi01,dia08,ima11}, and the
measured emission surface brightness is within the range explained by
star formation, again requiring no significant contribution from an 
energetically important AGN \citep{soi01,ima11}. 
The Pa$\alpha$ (1.88 $\mu$m) emission from star-forming HII-regions  
is dominated by nuclear $\sim$3$''$ area \citep{alo01}, which 
suggests that most of starburst activity in NGC 1614 is concentrated
in the nuclear regions within $<$a few arcsec. 

Although ALMA is a very powerful tool for unveiling the molecular gas
distribution in detail, it is not sensitive to spatially extended
structure beyond the maximum scale, which is $\sim$6$''$ at $\sim$350 GHz
in ALMA Cycle 0.   
The spatially resolved, but intrinsically compact ($\sim$2$''$) nuclear
emission morphology and the known starburst-dominated nature make NGC
1614 an ideal target to investigate the spatial distribution and
dynamics of dense molecular gas in merging LIRG nuclei, as well as to
obtain a template of line flux ratios of dense gas tracers in
starburst-dominated galaxies, during ALMA Cycle 0. 

We thus performed ALMA band 7 (275--373 GHz) observations of NGC
1614 to study the emission properties of HCN J = 4--3, HCO$^{+}$ J =
4--3, and HNC J = 4--3 lines.  
The basic information of NGC 1614 is summarized in Table 1. 
Throughout this paper, we adopt H$_{0}$ $=$ 71 km s$^{-1}$ Mpc$^{-1}$, 
$\Omega_{\rm M}$ = 0.27, and $\Omega_{\rm \Lambda}$ = 0.73 \citep{kom09}, 
where 1$''$ corresponds to $\sim$320 pc at the distance of NGC 1614 (z =
0.016). 

\section{Observations and Data Analysis}

All observations were made during ALMA Cycle 0 within the program 
2011.0.00020.S (PI = M. Imanishi).  
Observation details are described in Table 2. 
We adopted the widest 1.875 GHz band mode, and the total channel number
was 3840.

For NGC 1614 (z = 0.016), HCN J = 4--3 ($\nu_{\rm rest}$ = 354.505 GHz),
and HCO$^{+}$ J = 4--3 ($\nu_{\rm rest}$ = 356.734 GHz) lines are
simultaneously observable in ALMA band 7. 
Four frequency setups can cover four different frequencies at the same
time. 
Two were used to observe HCN J = 4--3 (central frequency was set as 
$\nu_{\rm center}$= 348.922 GHz) and HCO$^{+}$ J = 4--3 lines 
($\nu_{\rm center}$ = 350.920 GHz), and the remaining two were used
to measure the continuum flux level ($\nu_{\rm center}$ = 337.106 GHz and
338.681 GHz).  
The net on-source exposure time for the HCN/HCO$^{+}$ J = 4--3 line
observation of NGC 1614 was $\sim$26 min.

The frequency of the HNC J = 4--3 line ($\nu_{\rm rest}$ = 362.630 GHz) is 
separated from the HCN J = 4--3 and HCO$^{+}$ J = 4--3 lines, and 
we required independent observations.
The HNC line was covered with one spectral window 
($\nu_{\rm center}$ = 356.920 GHz), and an additional second spectral
window was used to probe the continuum emission 
($\nu_{\rm center}$ = 345.079 GHz).   
The net on-source exposure time for the HNC J = 4--3 line observation of
NGC 1614 was $\sim$25 min.   

We started data analysis from calibrated data provided by the Joint ALMA 
Observatory. 
We first checked the visibility plots to see if the signatures of the emission lines were recognizable. 
The presence of HCN, HCO$^{+}$, and HNC J = 4--3 lines were evident in the
visibility plot, whereas signatures of other molecular lines were not
clearly seen. 
We then selected channels that were free from strong line emission to
estimate the continuum flux level.
We subtracted this continuum level and performed the task ``clean'' for
molecular emission lines. 
The ``clean'' procedure was applied also to the continuum data. 
We employed 40-channel spectral binning ($\sim$17 km s$^{-1}$) and 0.3$''$
pixel$^{-1}$ spatial binning in this clean procedure. 

\section{Results}

Continuum emission properties are shown in Figure 1 and Table 3.
Figure 2 displays the integrated intensity (moment 0) maps of molecular
lines and spectra within the beam size at the continuum peak positions.
Continuum emission is found to be well subtracted in the spectra 
(Figure 2 right panels), and so the moment 0 maps in Figure 2 (left) 
should reflect the properties of individual molecular gas emission lines. 
The peak flux, rms noise level, and synthesized beams in individual maps
are summarized in Table 4 (row denoted ``all'').   
The peak positions of HCN J = 4--3 and HCO$^{+}$ J = 4--3 agree with that of
continuum ``a'', whereas that of HNC J = 4--3 is two pixels (0.6$''$)
displaced to the south from the continuum "b" peak.

CS J = 7--6 line at $\nu_{\rm rest}$ = 342.883 GHz is covered in a
spectral window during the observations of HCN/HCO$^{+}$ J = 4--3 line,
but it is not clearly detected in the spectrum (within the beam size)  
at the nucleus, which is defined from the continuum ``a'' emission peak
(Figure 2, right). 
Because the signals of the CS J = 7--6 emission are not clearly seen, 
to create the moment 0 map of CS J = 7--6, we refer
to the velocity profile of the brightest HCO$^{+}$ J = 4--3 emission line
at the nucleus and sum signals with v$_{\rm opt}$ 
$\equiv$ c ($\lambda$-$\lambda_{\rm 0}$)/$\lambda_{\rm 0}$ = 
4600--4925 [km s$^{-1}$]. 
No clear CS J = 7--6 emission line is seen ($<$3$\sigma$) in the
moment 0 map (Figure 2, left and Table 4). 
This undetected CS J = 7--6 line will not be used in the following
discussion.  

The moment 0 maps in Figure 2 display spatially extended structures
compared with the beam size, particularly for HCN and HCO$^{+}$ J = 4--3
lines.   
We created spectra of the HCN, HCO$^{+}$, and HNC J = 4--3 lines,
integrated over all regions with significant signal detection
($\sim$3$''$ $\times$ 3$''$), in Figure 3. 
All show double-peaked emission profiles, as seen in the $^{12}$CO
J = 2--1, $^{12}$CO J = 3--2, 
and $^{13}$CO J = 2--1 lines \citep{wil08}.
We applied double Gaussian fits. 
The derived parameters are summarized in Table 5 (rows denoted with ``all''). 
For HCO$^{+}$ J = 4--3 emission, a triple Gaussian fit was also applied
because three emission peaks were seen.
The molecular line luminosities, integrated over regions of
significant signal detection (corresponding to ``all'' in
Table 5) are summarized in Table 6, where we adopted equations (2) and
(3) of \citet{sol05}.

Previously obtained high-spatial-resolution images of starburst
indicators, such as the Pa$\alpha$ emission line (1.88 $\mu$m), the PAH
emission feature (3.3 $\mu$m), the infrared 8.7 $\mu$m continuum, and
the radio 5--8.4 GHz (3.6--6 cm) continuum, have revealed the presence
of ring-shaped circumnuclear starbursts with radii of 0.5--1.0$"$
\citep{nef90,alo01,dia08,ols10,vai12}. 
A similar ring pattern is
discernible in the high-spatial-resolution (0.5$"$ $\times$ 0.4$"$) CO
J = 2--1 molecular line map \citep{kon13}. 
Our ALMA map has a resolution of 1.5$"$ $\times$ 1.3$"$. 
In this map, particularly in the brightest HCO$^{+}$ J = 4--3 line,
where the highest S/N ratios are achieved, the emission is more
elongated toward the north-south direction than the east-west direction,
suggesting that a larger amount of high-density molecular gas is
distributed at the northern and southern part of the nucleus. 

Figure 4 displays the intensity weighted mean velocity (moment 1) and
intensity weighted velocity dispersion (moment 2) maps of HCN and
HCO$^{+}$ J = 4--3 lines.  
The dense molecular gas in the northern (southern)
part of the nucleus is clearly redshifted (blueshifted) relative to the 
nuclear velocity of NGC 1614 
(v$_{\rm opt}$ = 4800 [km s$^{-1}$] for z = 0.016). 
The contours of emission of the red component with v$_{\rm opt}$  
$>$ 4800 [km s$^{-1}$] and the blue component with v$_{\rm opt}$ $<$ 4800 
[km s$^{-1}$] are displayed in Figure 5.
The emission properties of the red and blue components are summarized in
Table 4 (rows denoted with ``red'' and ``blue'').
The emission peak positions of the red and blue components are shifted
by two pixels (0.6$''$) to the north and south direction,
respectively, from the nucleus. 
For HNC J = 4--3 emission in Figure 5, the blue component is much brighter
than the red component, which could explain the slight peak offset in
the HNC J = 4--3 moment 0 map (integrated over all velocity components with
significant signal detection) with respect to the nuclear position
(Table 3). 

Spectra within the beam size, at the nuclear position and the peak
position of the red and blue components, are shown in Figure 6.   
As expected, spectra at the red (blue) component peaks are redshifted
(blueshifted), relative to those at the nuclear position for all the
HCN, HCO$^{+}$, and HNC J = 4--3 lines.  
Gaussian fits were applied to the spectra.
The resulting parameters
are shown in Table 5 (rows denoted with ``peak'', ``red'', and ``blue'').
The line widths of these dense molecular gas tracers are generally
larger in the blue component than the red component, which has also been
seen in ionized gas \citep{der88}. 
This indicates that the southern part of the nucleus is more turbulent
than the northern part.

Figure 7 presents the position-velocity diagram along the north-south
direction, passing through the nucleus.
The channel map of the brightest HCO$^{+}$ J = 4--3 emission line is 
shown in Figure 8. 
The dense molecular gas in the northern region has greater velocity
(more redshifted) than the southern molecular gas.  

\section{Discussion}

\subsection{Spatial distribution of dense molecular gas}

The red and blue components of the dense molecular gas show emission
peaks at $\sim$0.6$''$ north and south of the nucleus, respectively,
roughly corresponding to the northern and southern parts of the
previously identified circumnuclear starburst ring with a radius of 
0.5--1$''$ \citep{nef90,alo01,dia08,ols10,vai12}.
\citet{vai12} investigated the spatial distribution of the 3.3 $\mu$m 
PAH emission feature, and found particularly strong PAH emission at the
northern and southern parts of the ring. 
The 3.3 $\mu$m PAH emission is a good indicator of starburst
activity \citep{moo86,imd00,ima06a,ima08,ima10b}.

In contrast, the Pa$\alpha$ (1.88 $\mu$m) emission, originating in
HII regions, is strongest at the eastern and western sides of the
circumnuclear starburst ring \citep{alo01}. 
How are these two kinds of starbursts related?
One scenario involves an age difference.
The eastern and western starburst regions are probed with the tracers of  
HII regions where plenty of ionizing ($\lambda$ $<$ 912 $\rm \AA$) 
UV photons, usually dominated by short-lived massive O-stars, are needed. 
In contrast, the 3.3 $\mu$m PAH emission mostly comes from
photo-dissociation regions between HII regions and molecular gas, where
PAHs are excited by non-ionizing ($\lambda$ $>$ 912 $\rm \AA$) stellar
UV photons \citep{sel81}.  
For the PAH-exciting non-ionizing UV continuum, the contribution  
from less massive stars than O stars is higher than that for ionizing UV
photons inside HII regions. Thus, the PAH emission feature is sensitive
not only to very young O-stars dominant starbursts but also to slightly 
aged starbursts where massive O-stars have mostly died, but less massive
stars (e.g., B-stars) still survive and emit a sufficient quantity of
non-ionizing PAH-exciting UV photons. 
\citet{dia08} showed that the 8.7 $\mu$m infrared dust continuum,
relative to the HII region tracer Pa$\alpha$ line (1.88 $\mu$m), is
enhanced at the northern and southern starburst ring.
Because non-ionizing UV continuum emission can contribute significantly to
the infrared continuum emission but cannot do so for the Pa$\alpha$ emission,
the observed spatial variation in the 3.3 $\mu$m PAH to Pa$\alpha$ and 
8.7 $\mu$m to Pa$\alpha$ flux ratio is explainable under the scenario
that the typical starburst age is older at the northern and southern
ring than at the eastern and western ring \citep{dia08}. 
If the circumnuclear starburst ring is formed by nuclear starbursts
that are progressing outward \citep{alo01}, the age difference in 
starbursts at different positions of the ring needs to be explained.

Dust extinction is another possibility because Pa$\alpha$ emission at
1.88 $\mu$m can be more highly flux-attenuated than the 3.3 $\mu$m PAH
emission and the 8.7 $\mu$m dust continuum. 
Radio free-free continuum emission from HII regions is less susceptible
to foreground dust extinction and could help to determine whether Pa$\alpha$
emission is significantly affected by dust extinction. 
High-spatial-resolution radio continuum maps at 5 GHz (6 cm) and 8.4 GHz
(3.6 cm) are available \citep{nef90,ols10} and \citet{ols10} ascribe
the radio 5 GHz and 8.4 GHz emission in NGC 1614 to free-free emission
from HII-regions in young starbursts.   
Because the effects of free-free absorption inside HII regions are smaller
at 8.4 GHz than at 5 GHz, the 8.4 GHz radio emission map is taken as the
better probe of the true spatial distribution of HII regions.  

In the radio 8.4 GHz map, strong emission is detected in the
eastern and western starburst ring \citep{ols10,kon13}, as seen in
Pa$\alpha$ line map \citep{alo01}, confirming that luminous HII regions
are present at those locations. 
However, at 8.4 GHz, bright emission is seen also at the northern
starburst ring, where Pa$\alpha$ emission is not strong \citep{kon13}.  
Similarly, when compared with the eastern starburst ring, 
the southern starburst ring is more conspicuous at 8.4 GHz than in
Pa$\alpha$ \citep{kon13}.  
The comparison of the radio 8.4 GHz (3.6 cm) and 1.88 $\mu$m Pa$\alpha$
emission indicates that HII-regions at the northern and southern parts
of the starburst ring, unveiled by the radio 8.4 GHz emission, are not
sufficiently distinguished by Pa$\alpha$ emission. 
These northern and southern regions of the starburst ring are the
locations where dense molecular gas is distributed, according to our ALMA
data. 
Given that dust coexists with dense molecular gas, dust extinction is a 
natural explanation for the small Pa$\alpha$ to 8.4 GHz flux ratio at the
northern and southern starburst rings.  
Weak dust extinction is reported for starburst regions in NGC 1614 
compared with other general starburst galaxies \citep{alo01}, based
on near-infrared observations at $\lambda$ $<$ 2 $\mu$m.
This could be due partly to the fact that observations at $\lambda$ $<$ 2
$\mu$m, including Pa$\alpha$ emission, selectively trace emission from
less dusty starburst regions at the eastern and western ring and do not
properly probe the dusty starbursts at the northern and southern 
parts of the ring due to flux attenuation by dust extinction.  

\subsection{Dynamics of dense molecular gas}

Our ALMA data show that the high-density molecular gas in the northern part
of the nucleus is redshifted and that gas in the southern part is
blueshifted with respect to the nuclear velocity of this galaxy.
This suggests the rotation of dense molecular gas along the east-west
axis \citep{nef90}.   
A similar velocity pattern was found previously in the ionized gas maps
\citep{der88} and the lower density molecular gas probed with  
CO J = 3--2 and J = 2--1 \citep{wil08,kon13}. 
The observed velocity dispersion is highest in the nuclear region with
$\sim$80 km s$^{-1}$ (Figure 4), although it may be affected by 
the beam smearing of a rotating motion at the center.
This value is similar to those measured through near-infrared 
spectroscopy \citep{shi94} and with (sub)millimeter 
CO J = 3--2 and CO J = 2--1 emission lines \citep{wil08}.

The rotational motion found in our moment 1 maps (Figure 4) can be used
to derive the dynamical mass inside the rotating dense molecular gas
disk.  
We used the HCO$^{+}$ J = 4--3 line because it is brighter than HCN J =
4--3, and so the achieved S/N ratios are higher. 
In Figure 4, the rotational velocity is v $\sim$ 100 km s$^{-1}$ at
1.5--2$''$ (or r = 480--640 pc at z = 0.016) from the nucleus. 
The derived dynamical mass within 1.5--2$''$ radius is M$_{\rm dyn}$ =
rv$^{2}$/G/sin(i)$^{2}$ = 1.5--2.5 $\times$ 10$^{9}$M$_{\odot}$, where
the inclination angle i = 51$^{\circ}$ is adopted \citep{der88,alo01}. 
This mass is comparable to the previously estimated values
\citep{shi94,alo01,ols10}.  

\subsection{Flux ratios of dense gas tracers}

Figure 9 is a plot of HCN-to-HCO$^{+}$ and HCN-to-HNC flux ratios at 
the J = 4--3 transition, derived from the spectra at the nucleus, red
component peak position, blue component peak, and all regions with
significant signal detection. 
In all data, HCN J = 4--3 flux is smaller than that of HCO$^{+}$
J = 4--3, but higher than HNC J = 4--3 flux.

Using the J = 1--0 transitions of HCN, HCO$^{+}$, and HNC, the possibility 
of distinguishing the hidden energy sources of merging LIRG's dusty
nuclei is suggested \citep{koh05,ima04,ima06b,ima07,per07,kri08,ima09,cos11}. 
In general, HCN-to-HCO$^{+}$ flux ratios are small ($<$1), and HCN-to-HNC 
flux ratios are large ($>$1) in starburst-dominated galaxies, whereas 
HCN-to-HCO$^{+}$ flux ratios can be high ($>$1) in AGNs.
AGNs could enhance HCN flux relative to HCO$^{+}$, due possibly to HCN
abundance enhancement by AGN radiation \citep{har10} and/or infrared
radiative pumping of HCN \citep{sak10}. 
The low HCN-to-HCO$^{+}$ flux ratios obtained in the starburst-dominated
galaxy NGC 1614 at J = 4--3 are similar to other starburst galaxies at J
= 1--0. 
To obtain a physical interpretation of the observed J = 4--3 flux ratio,
we need to know the excitation conditions. 

The HCN J = 1--0 flux of NGC 1614 was found to be 7.2 [Jy km s$^{-1}$]
by \citet{gao04} based on single dish telescope observations. 
Under thermal excitation, the HCN J = 4--3 flux is expected to be 
16 times higher than J = 1--0, and so $\sim$115 [Jy km s$^{-1}$].
Our ALMA observations provide an observed HCN J = 4--3 flux from all
signal-detected regions of 2.8 [Jy km s$^{-1}$].
The smaller flux of our ALMA data could partly be caused by missing
flux, as our ALMA observations are insensitive to spatially extended
emission with $>$6$''$.  
\citet{sco89} and \citet{wil08} estimated that in NGC 1614, the nuclear
($<$several arcsec) CO J = 1--0 and J = 3--2 emission can account for
$>$30\% and $>$45\% of the total flux measured with single-dish telescopes.
Because the HCN J = 4--3 line traces higher density molecular gas
($>$10$^{6}$ cm$^{-3}$) than do CO J = 1--0 and J = 3--2 lines and because
higher-density gas is more concentrated in the nuclear region, our ALMA
HCN J = 4--3 data should recover $>$45\% of the total flux.
Even assuming a missing flux of a factor of $\sim$2, the HCN J = 4--3 flux  
of NGC 1614 is 5.6 [Jy km s$^{-1}$], only $<$5\% of the expected 
flux (115 Jy km s$^{-1}$) for thermal excitation.
Thus, HCN J = 4--3 line is significantly sub-thermally excited in NGC
1614, as observed in nearby galaxies at $>$100 pc scale \citep{knu07}.

Since the critical density of HCN J = 4--3 (n$_{\rm crit}$ $\sim$ 
2 $\times$ 10$^{7}$ cm$^{-3}$) is higher than that of HCO$^{+}$ J = 4--3 
(n$_{\rm crit}$ $\sim$ 4 $\times$ 10$^{6}$ cm$^{-3}$) \citep{mei07},
HCO$^{+}$ J = 4--3 is more easily excited than HCN J = 4--3 in starbursts.
In an AGN, the emission surface brightness is higher than starburst
activity, so the surrounding dust and gas are heated to a higher
temperature, 
which may help to excite HCN J = 4--3 more than starburst activity. 
A high HCN-to-HCO$^{+}$ J = 4--3 flux ratio could be a good diagnostic of
an AGN, simply because of the high excitation of HCN J = 4--3 in an AGN,
even without an HCN-abundance enhancement \citep{har10}. 
HCN-to-HCO$^{+}$ flux ratios of known AGN-important galaxies are being
measured in our ALMA program (Imanishi et al., in preparation), and they
tend to show higher HCN-to-HCO$^{+}$ J = 4--3 flux ratios than NGC 1614,
the template starburst galaxy (see also Imanishi et al. 2010a; Sakamoto
et al. 2010; Iono et al. 2013). 
Since the J = 4--3 lines of HCN and HCO$^{+}$ are at higher frequencies
(shorter wavelengths) than the lower J transition lines, the empirical
energy diagnostic method, if established at J = 4--3, is applicable to
more distant merging LIRGs using ALMA.    
This advantage is strengthened if HCN excitation is generally thermal
up to the J = 4--3 transition in AGN-important galaxies because
HCN flux increases proportional to the square of frequency, partly
compensating for the increase in Earth's atmospheric background emission
at higher-frequency ALMA bands.   
However, if the excitation at HCN J = 4--3 is sub-thermal, then HCN J =
3--2 or J = 2--1 lines may be better tracers of AGN in terms of actually
obtainable S/N ratios with ALMA. 
Additional molecular line transition data at J = 3--2 and 2--1 for
starburst-dominated galaxies and AGN-important galaxies are needed (1)
to distinguish whether high HCN-to-HCO$^{+}$ flux ratios at 
J = 4--3 in AGN-important galaxies are due to HCN abundance enhancement
and/or more HCN J = 4--3 excitation than starbursts and (2) to identify
the J transition lines that  are practically the best diagnostic for
separating AGNs from starburst-dominated galaxies. 

\section{Summary}

We performed HCN, HCO$^{+}$, and HNC J = 4--3 line observations of the
well-studied starburst galaxy NGC 1614 to trace the properties of the
high-density molecular gas. 
Our results are summarized as follows:

\begin{enumerate}
\item HCN, HCO$^{+}$, and HNC J = 4--3 emission are clearly detected in 
the nuclear regions of NGC 1614, but CS J = 7--6 emission and HCN J=4--3 
line at a vibrationally-excited level (v$_{2}$=1, l=1f) are not. 

\item HCN, HCO$^{+}$, and HNC J = 4--3 emission at the northern and southern 
parts of the nucleus are redshifted and blueshifted, respectively, with
respect to the nuclear velocity of this galaxy.
When the emission is separated into the red and blue components, the red
and blue components are strongest at 0.6$''$ north and south of the
nucleus for all of the HCN, HCO$^{+}$, and HNC J = 4--3 lines.   

\item At the peak location of the red and blue components of these 
dense molecular gas tracers, the presence of active dusty starbursts is
suggested, based on the infrared 3.3 $\mu$m PAH emission, infrared 
8.7 $\mu$m dust continuum emission, and radio 8.4 GHz free-free
emission, supporting the scenario that starbursts occur in dense
molecular gas.  

\item The dynamical mass derived from the red and blue dense molecular
gas components, assuming rotational motion, is 1.5--2.5 $\times$
10$^{9}$M$_{\odot}$ within $\sim$2$''$ in radius.
This is similar to estimates previously obtained using other methods. 

\item The HCN-to-HCO$^{+}$ flux ratios are smaller than unity, and the 
HCN-to-HNC flux ratios are higher than unity for J = 4--3 in NGC 1614, 
which is a similar trend to previous observations of J = 1--0 for
starburst-dominated galaxies.   
\end{enumerate}

\acknowledgments

We thank E. Mullar and H. Nagai for their useful advice on ALMA data
analysis. 
M.I. is supported by Grants-in-Aid for Scientific Research (no. 22012006).  
This paper makes use of the following ALMA data:
ADS/JAO.ALMA\#2011.0.00020.S . ALMA is a partnership of ESO
(representing its member states), NSF (USA), and NINS (Japan), together
with NRC (Canada) and NSC and ASIAA (Taiwan), in cooperation with the
Republic of Chile. The Joint ALMA Observatory is operated by ESO,
AUI/NRAO, and NAOJ. 

\clearpage


\clearpage

\begin{deluxetable}{lcrrrrc}
\tabletypesize{\small}
\tablecaption{The {\it IRAS}-based infrared emission properties of 
NGC 1614 \label{tbl-1}}
\tablewidth{0pt}
\tablehead{
\colhead{Object} & \colhead{Redshift}   & 
\colhead{f$_{\rm 12}$}   & 
\colhead{f$_{\rm 25}$}   & 
\colhead{f$_{\rm 60}$}   & 
\colhead{f$_{\rm 100}$}  & 
\colhead{log L$_{\rm IR}$}  \\
\colhead{} & \colhead{}   & \colhead{[Jy]} & \colhead{[Jy]} 
& \colhead{[Jy]} & \colhead{[Jy]}  & \colhead{[L$_{\odot}$]} \\
\colhead{(1)} & \colhead{(2)} & \colhead{(3)} & \colhead{(4)} & 
\colhead{(5)} & \colhead{(6)} & \colhead{(7)} 
}
\startdata
NGC 1614 (IRAS 04315$-$0840) & 0.016 & 1.38 & 7.50 & 32.12 & 34.32 & 11.6 \\  
\enddata

\tablecomments{
Col.(1): Object name. 
Col.(2): Redshift. 
Col.(3)--(6): f$_{12}$, f$_{25}$, f$_{60}$, and f$_{100}$ are 
{\it IRAS} fluxes at 12 $\mu$m, 25 $\mu$m, 60 $\mu$m, and 100 $\mu$m,
respectively, taken from \citet{san03}. 
Col.(7): Decimal logarithm of infrared (8$-$1000 $\mu$m) luminosity
in units of solar luminosity (L$_{\odot}$), calculated with
$L_{\rm IR} = 2.1 \times 10^{39} \times$ D(Mpc)$^{2}$
$\times$ (13.48 $\times$ $f_{12}$ + 5.16 $\times$ $f_{25}$ +
$2.58 \times f_{60} + f_{100}$) [ergs s$^{-1}$] \citep{sam96}.
}

\end{deluxetable}

\begin{deluxetable}{llcccc}
\tabletypesize{\small}
\tablecaption{Log of ALMA Cycle 0 observations of NGC 1614 \label{tbl-2}}
\tablewidth{0pt}
\tablehead{
\colhead{Line} & \colhead{Date} & \colhead{Antenna} & 
\multicolumn{3}{c}{Calibrator} \\ 
\colhead{} & \colhead{[UT]} & \colhead{Number} &
\colhead{Bandpass} & \colhead{Flux} & \colhead{Phase}  \\
\colhead{(1)} & \colhead{(2)} & \colhead{(3)} & \colhead{(4)} &
\colhead{(5)} & \colhead{(6)}   
}
\startdata 
HCN/HCO$^{+}$ J=4--3 & 2011 November 15 & 16 & 3C454.3 & Callisto & J0423$-$013\\
HNC J=4--3 & 2011 November 15 & 16 & 3C454.3 & Callisto & J0423$-$013\\
\enddata

\tablecomments{
Col.(1): Observed molecular line. 
Col.(2): Observing date in UT. 
Col.(3): Number of antennas used for observations.
Cols.(4), (5), and (6): Bandpass, flux, and phase calibrators used
for our NGC 1614 observations, respectively.   
}

\end{deluxetable}

\begin{deluxetable}{ccrccl}
\tabletypesize{\scriptsize}
\tablecaption{Continuum emission of NGC 1614 \label{tbl-3}}
\tablewidth{0pt}
\tablehead{
\colhead{Continuum} & \colhead{Frequency} & \colhead{Flux} & 
\colhead{Peak Coordinate} & \colhead{rms} & \colhead{Synthesized Beam} \\
\colhead{} & \colhead{[GHz]} & \colhead{[mJy beam$^{-1}$]} & 
\colhead{(RA,DEC)} & \colhead{[mJy beam$^{-1}$]} & 
\colhead{[arcsec $\times$ arcsec] ($^{\circ}$)} \\  
\colhead{(1)} & \colhead{(2)} & \colhead{(3)}  & \colhead{(4)}  &
\colhead{(5)} & \colhead{(6)}  
}
\startdata 
a & 344 & 14.6 (45$\sigma$) \tablenotemark{a} & (04 34 00.01, $-$08 34 44.9) & 0.32 & 1.5
$\times$ 1.3 (81$^{\circ}$) \\ 
b & 351 & 10.9 (29$\sigma$) \tablenotemark{a} & (04 34 00.03, $-$08 34 44.9) & 0.37 & 1.3
$\times$ 1.3 ($-$53$^{\circ}$)\\   
\enddata

\tablenotetext{a}{We do not include possible systematic uncertainty, which is difficult 
to quantify.}

\tablecomments{
Col.(1): Continuum "a" and "b" data were taken during observations
of HCN/HCO$^{+}$ J = 4--3 and HNC J = 4--3, respectively.
Col.(2): Central frequency of the continuum in [GHz].
Col.(3): Peak signal value in the continuum map in [mJy beam$^{-1}$],
and detection significance, relative to the rms noise, in parentheses.
Col.(4): The coordinate of the continuum emission peak in J2000.
Col.(5): The rms noise (1$\sigma$) of the continuum map in [mJy beam$^{-1}$].
Col.(6): The synthesized beam of the continuum map.
Position angle is 0$^{\circ}$ along the north-south direction, and
increases in the counter-clockwise direction. 
}

\end{deluxetable}

\clearpage

\begin{deluxetable}{lcccl}
\tabletypesize{\small}
\tablecaption{Molecular line flux in NGC 1614 \label{tbl-4}}
\tablewidth{0pt}
\tablehead{
\colhead{Line} & \colhead{Velocity} & \multicolumn{3}{c}{Integrated 
  intensity (moment 0) map} \\
\colhead{} & \colhead{component} & \colhead{Peak} &
\colhead{rms} & \colhead{Beam} \\
\colhead{}& \colhead{} & \colhead{[Jy beam$^{-1}$ km s$^{-1}$]} &
\colhead{[Jy beam$^{-1}$ km s$^{-1}$]} & 
\colhead{[arcsec $\times$ arcsec] ($^{\circ}$)}\\
\colhead{(1)} & \colhead{(2)} & \colhead{(3)} & \colhead{(4)} & 
\colhead{(5)}  
}
\startdata 
HCN J=4--3 & all & 1.7 (12$\sigma$) & 0.14 & 1.5 $\times$
1.3 (76$^{\circ}$) \\
           & red & 1.3 (14$\sigma$) & 0.088 & 1.5 $\times$
1.3 (76$^{\circ}$) \\
           & blue & 0.99 (9.4$\sigma$) & 0.11 & 1.5 $\times$ 1.3
(76$^{\circ}$) \\ \hline 
 HCO$^{+}$ J=4--3 & all & 7.1 (33$\sigma$) & 0.22 & 1.5 $\times$
1.3 (76$^{\circ}$) \\  
           & red & 5.2 (42$\sigma$) & 0.12 & 1.5 $\times$ 1.3
(76$^{\circ}$) \\ 
           & blue & 5.7 (36$\sigma$) & 0.16 & 1.5 $\times$ 1.3
(76$^{\circ}$) \\ \hline 
HNC J=4--3 & all & 1.1 (7.0$\sigma$) \tablenotemark{a} & 0.16 & 1.3
$\times$ 1.3 ($-$53$^{\circ}$) \\   
           & red & 0.57 (6.9$\sigma$) & 0.084 & 1.3 $\times$ 1.3
($-$53$^{\circ}$) \\ 
           & blue & 1.1 (8.2$\sigma$) & 0.13 & 1.3 $\times$ 1.3
($-$54$^{\circ}$) \\ \hline 
CS J=7--6 & all & $<$0.45($<$3$\sigma$)  \tablenotemark{b} & 0.15 & 1.5
$\times$ 1.4 (81$^{\circ}$) \\
\enddata

\tablenotetext{a}{The peak position of the HNC J = 4--3 emission,
integrating over all velocity components with significant signal detection,
is two pixels (0.6$''$) south of the continuum ``b'' peak, which could be
explained by the stronger blue HNC emission component compared with the red
component (Figure 5). See text in $\S$3.} 

\tablenotetext{b}{The HCN J=4--3 to CS J=7--6 flux ratio is $>$3.7.
This lower limit is lower than the ratios found in AGNs, and 
is comparable to those observed in starburst galaxies \citep{izu13}.
}

\tablecomments{
Col.(1): Molecular line.
Col.(2): Velocity component. 
The notations ``all'', ``red'', and ``blue'' mean 
all velocity components with significant signal detection, 
red component with v$_{\rm opt}$ 
$\equiv$  c ($\lambda$-$\lambda_{\rm 0}$)/$\lambda_{\rm 0}$
$>$ 4800 km s$^{-1}$, and
blue component with v$_{\rm opt}$ $<$ 4800 km s$^{-1}$, respectively.
Col.(3): Peak flux in the integrated intensity (moment 0) map in 
[Jy beam$^{-1}$ km s$^{-1}$] and detection significance relative to
the rms noise, in parentheses.
Col.(4): The rms noise level (1$\sigma) $ in the moment 0 map in 
[Jy beam$^{-1}$ km s$^{-1}$]. 
Col.(5): Synthesized beam of the moment 0 map.
Position angle is 0$^{\circ}$ along the north-south direction and
increases in the counter-clockwise direction. 
}

\end{deluxetable}

\begin{deluxetable}{ll|cccc}
\tabletypesize{\scriptsize}
\tablecaption{Gaussian fit parameters to molecular line emission 
from NGC 1614 \label{tbl-5}}
\tablewidth{0pt}
\tablehead{
\colhead{Line} & \colhead{Position} &
\multicolumn{4}{c}{Gaussian line fit} \\   
\colhead{} & \colhead{} & \colhead{Center} & \colhead{Peak}
& \colhead{FWHM} & \colhead{Flux} \\  
\colhead{} & \colhead{} & \colhead{[km s$^{-1}$]} &
\colhead{[mJy]} & \colhead{[km s$^{-1}$]} & \colhead{[Jy km s$^{-1}$]} \\ 
\colhead{(1)} & \colhead{(2)} & \colhead{(3)} & \colhead{(4)} & 
\colhead{(5)} & \colhead{(6)} 
}
\startdata 
HCN J=4--3 & all  & 4733$\pm$25 + 4857$\pm$14 & 11$\pm$2 +
12$\pm$4 & 171$\pm$52 + 73$\pm$30 & 2.8$\pm$0.8 \\
           & peak & 4822$\pm$10 & 11$\pm$1 & 131$\pm$24 & 1.6$\pm$0.3 \\
           & red & 4853$\pm$2 & 18$\pm$2 & 68$\pm$7 & 1.3$\pm$0.2 \\
           & blue & 4720$\pm$17 & 6.2$\pm$0.9 & 214$\pm$42 & 1.4$\pm$0.3 \\
HCO$^{+}$ J=4--3 & all & 4718$\pm$9 + 4858$\pm$3 & 50$\pm$3 +
70$\pm$6 & 200$\pm$20 + 72$\pm$8 & 16.0$\pm$1.4\tablenotemark{a} \\  
           &     & 4650$\pm$4 + 4755$\pm$5 + 4857$\pm$2 & 51$\pm$6 +
49$\pm$2 + 81$\pm$8 & 70$\pm$8 + 105$\pm$31 + 72$\pm$5 & 15.6$\pm$1.9\tablenotemark{a} \\ 
           & peak & 4803$\pm$4 & 41$\pm$2 & 167$\pm$9 & 7.3$\pm$0.5 \\
           & red & 4854$\pm$1 & 72$\pm$3 & 76$\pm$2 & 5.8$\pm$0.3 \\
           & blue & 4715$\pm$4 & 37$\pm$2 & 180$\pm$9 & 7.0$\pm$0.5 \\
HNC J=4--3 & all & 4718$\pm$11 + 4859$\pm$16 & 10$\pm$2 + 7.8$\pm$2.7
& 125$\pm$27 + 50$\pm$45 & 1.7$\pm$0.5 \\ 
           & peak & 4800 (fix) & 3.6$\pm$1.0 & 183$\pm$75 & 0.7$\pm$0.3 \\  
           & red  & 4856$\pm$3 & 13$\pm$2 & 42$\pm$7 & 0.6$\pm$0.1 \\
           & blue & 4714$\pm$9 & 9.1$\pm$1.4 & 116$\pm$23 & 1.1$\pm$0.3 \\
\enddata

\tablenotetext{a}{\citet{wil08} derived the flux to be $>$14$\pm$3 
[Jy km s$^{-1}$] based on the Submillimeter Array (SMA) data, which
probe emission over the spatial extent of $<$7$''$.}

\tablecomments{
Col.(1): Molecular line.
Col.(2): Position and area for spectral extraction.
``all'' means spectra integrated over all regions of significant
signal detection ($\sim$3$''$ $\times$ 3$''$).  
The terms ``peak'', ``red'', and ``blue'' denote spectra within the beam
size at the peak position of the continuum, red, and blue molecular line
components, respectively. 
The coordinates of the red and blue peaks are (04 34 00.01, $-$08 34 44.3)
and (04 34 00.01, $-$08 34 45.5) in J2000, respectively, for all of the HCN,
HCO$^{+}$, and HNC. 
For the ``peak'', the continuum ``a'' peak coordinate (Table 3) is used for
HCN and HCO$^{+}$, and the continuum ``b'' peak (Table 3) is used for HNC.
Cols. (3)-(6): Gaussian fits of the detected molecular emission lines.
For ``all'', double Gaussian fits are applied because emission lines
are double peaked.
For ``all''  of HCO$^{+}$, a triple Gaussian fit is also applied for
comparison. 
Col.(3): Central velocity of the Gaussian fits in [km s$^{-1}$].
Col.(4): Peak flux of the Gaussian fits in [mJy].
Col.(5): Full width at half maximum (FWHM) of the Gaussian fits in 
[km s$^{-1}$]. 
Col.(6): Total line flux, based on the Gaussian fits, in [Jy km s$^{-1}$].
}

\end{deluxetable}

\begin{deluxetable}{lcr}
\tablecaption{Molecular line luminosity for NGC 1614 \label{tbl-6}}
\tablewidth{0pt}
\tablehead{
\colhead{Line} & \colhead{[L$_{\odot}$]} &
\colhead{[K km s$^{-1}$ pc$^{2}$]} \\   
\colhead{(1)} & \colhead{(2)} & \colhead{(3)} 
}
\startdata 
HCN J=4--3 & (4.8$\pm$1.4)$\times$10$^{3}$ & (3.4$\pm$1.0)$\times$10$^{6}$ \\
HCO$^{+}$ J=4--3 & (2.7$\pm$0.2)$\times$10$^{4}$ & (1.9$\pm$0.2)$\times$10$^{7}$ \\
HNC J=4--3 & (3.0$\pm$0.9)$\times$10$^{3}$ & (1.9$\pm$0.6)$\times$10$^{6}$ \\
\enddata

\tablecomments{
Col.(1): Molecular line.
Col.(2): Luminosity in units of [L$_{\odot}$].
Col.(3): Luminosity in units of [K km s$^{-1}$ pc$^{2}$].
}

\end{deluxetable}

\clearpage

\begin{figure}
\begin{center}
\includegraphics[angle=0,scale=.5]{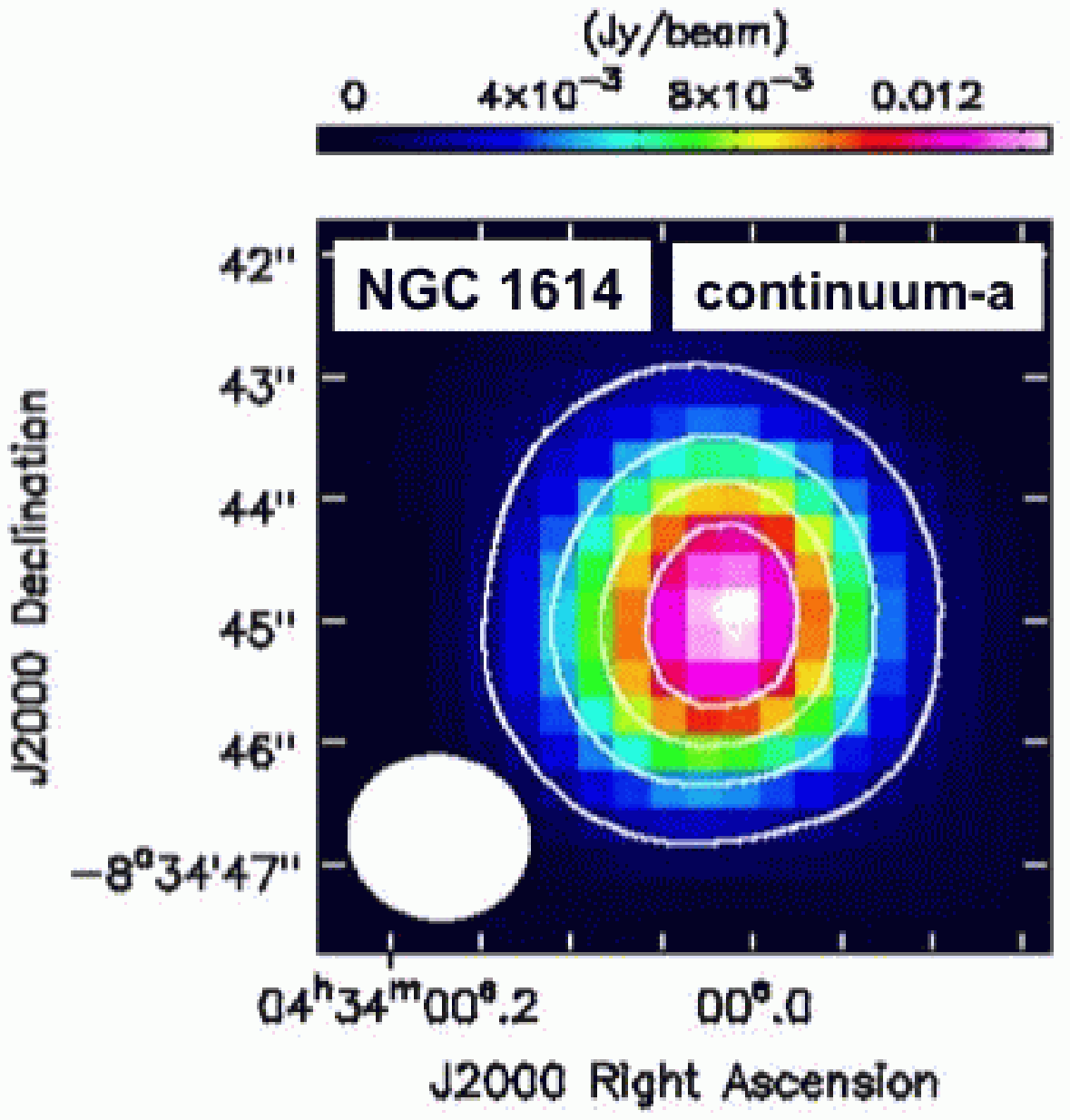} 
\includegraphics[angle=0,scale=.5]{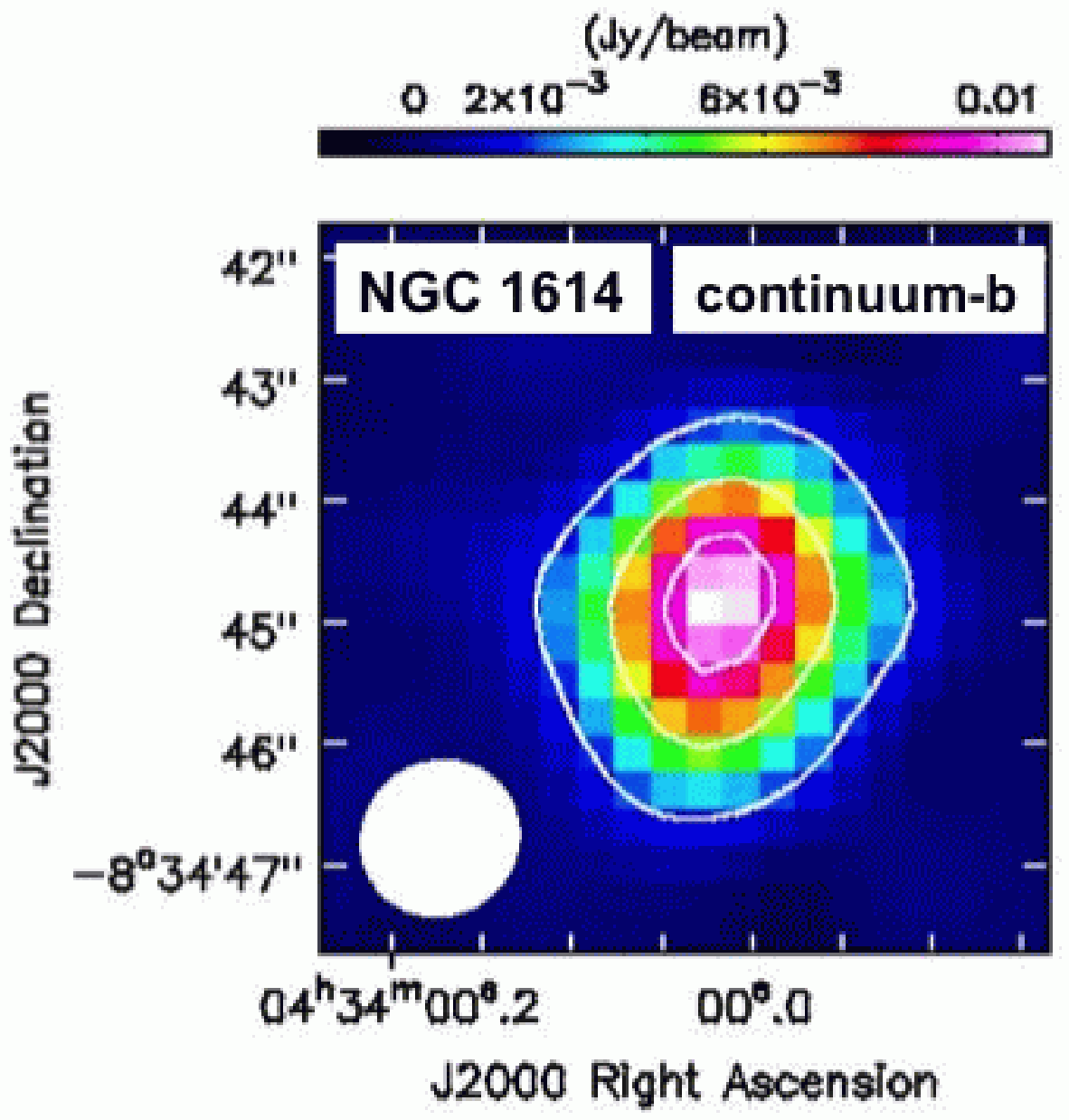} 
\end{center}
\caption{
Continuum maps. North is top, and east is to the left.
The continuum ``a''  ({\it Left}) and ``b''  ({\it Right}) data were taken
during observations of the HCN/HCO$^{+}$ J = 4--3 and HNC J = 4--3 lines,
respectively. 
Contours are 5$\sigma$, 15$\sigma$, 25$\sigma$, 35$\sigma$, 45$\sigma$
for continuum ``a'' and 5$\sigma$, 15$\sigma$, 25$\sigma$ for continuum
``b''.
The 1 $\sigma$ level is shown in Table 3 and slightly differs between
continua ``a'' and ``b''.  
The continuum ``a'' data have higher detection significance than the 
continuum ``b'' data, and the peak position of the continuum ``a'' 
emission is (RA, DEC) = (04 34 00.01, $-$08 34 44.9) in J2000. 
We define this coordinate as the nucleus of NGC 1614 in this paper. 
The synthesized beams are shown as filled white circles at the bottom
left of the individual figures.
}
\end{figure}

\begin{figure}
\begin{center}
\includegraphics[angle=0,scale=.45]{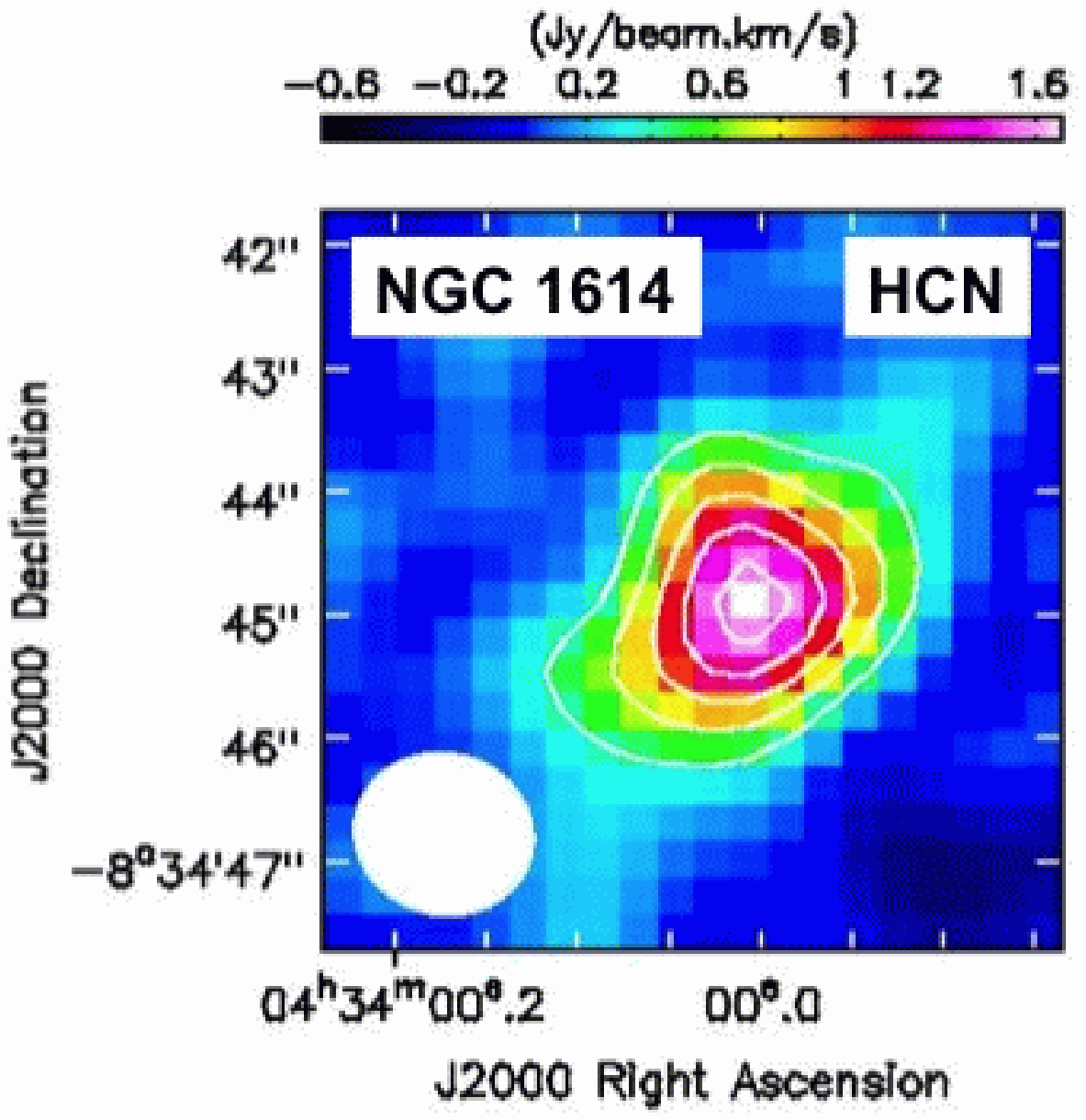} 
\includegraphics[angle=0,scale=.5]{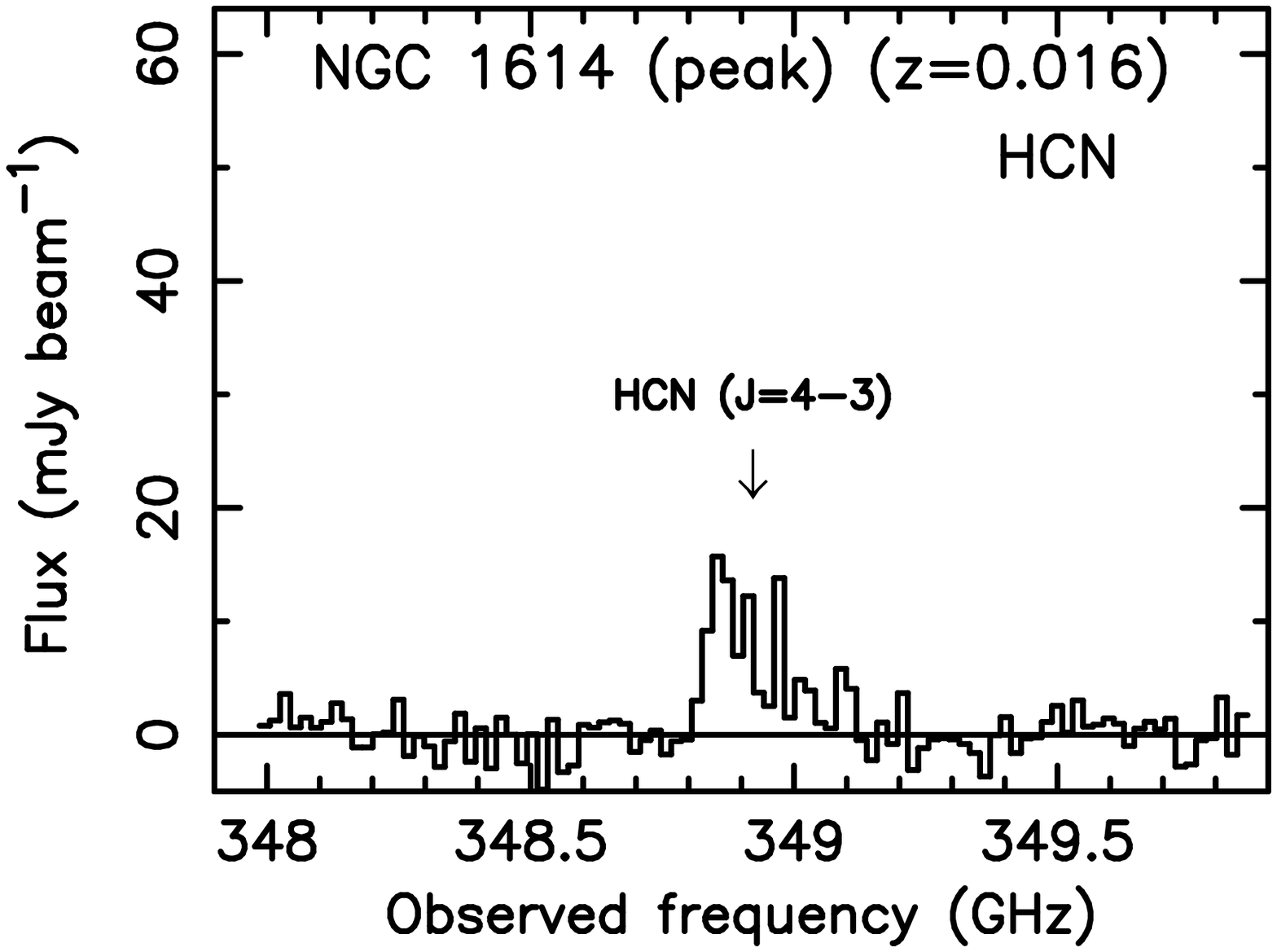}\\ 
\includegraphics[angle=0,scale=.45]{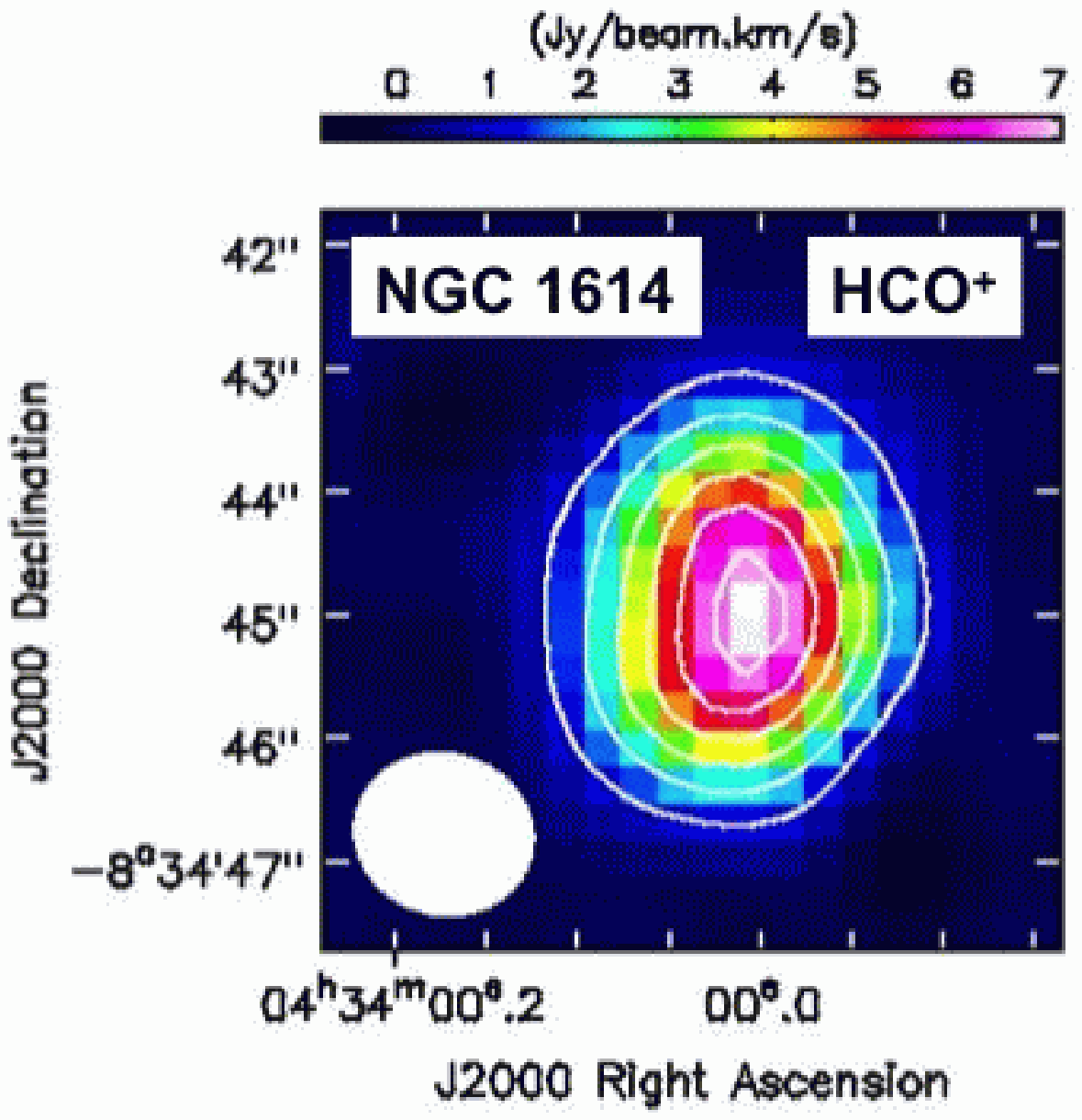} 
\includegraphics[angle=0,scale=.5]{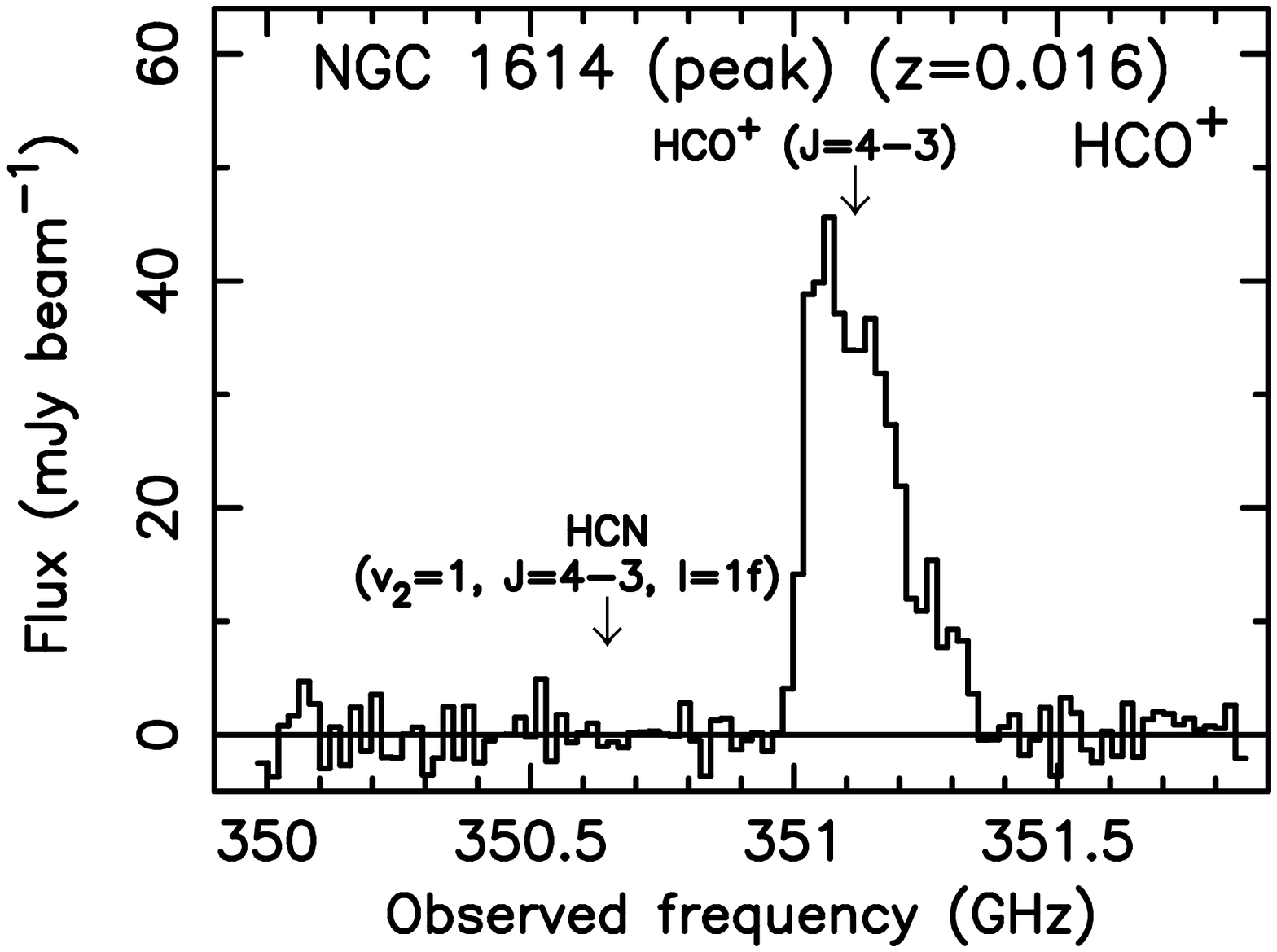}\\ 
\includegraphics[angle=0,scale=.45]{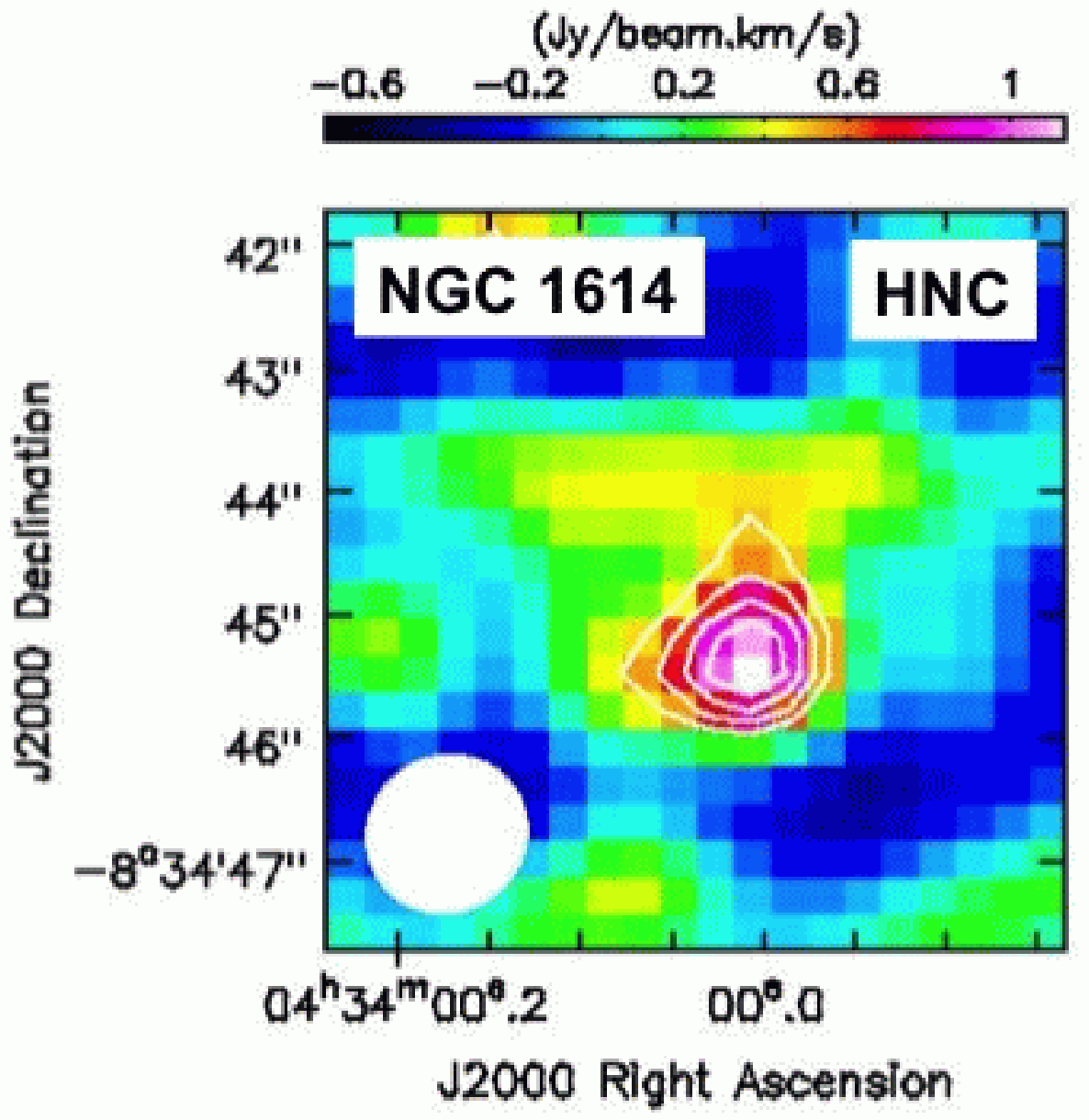} 
\includegraphics[angle=0,scale=.5]{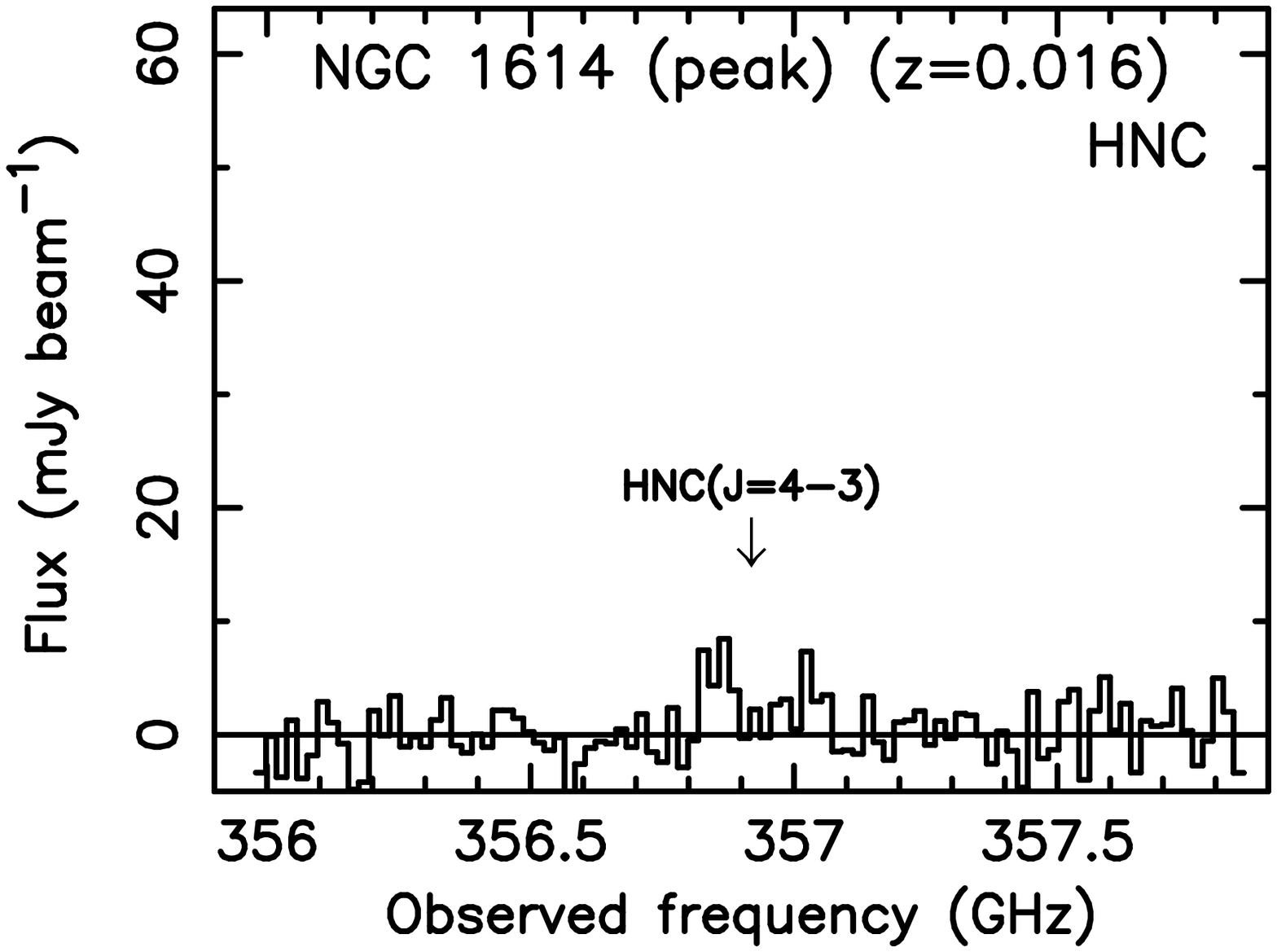}\\ 
\end{center}
\end{figure}

\begin{figure}
\begin{center}
\includegraphics[angle=0,scale=.45]{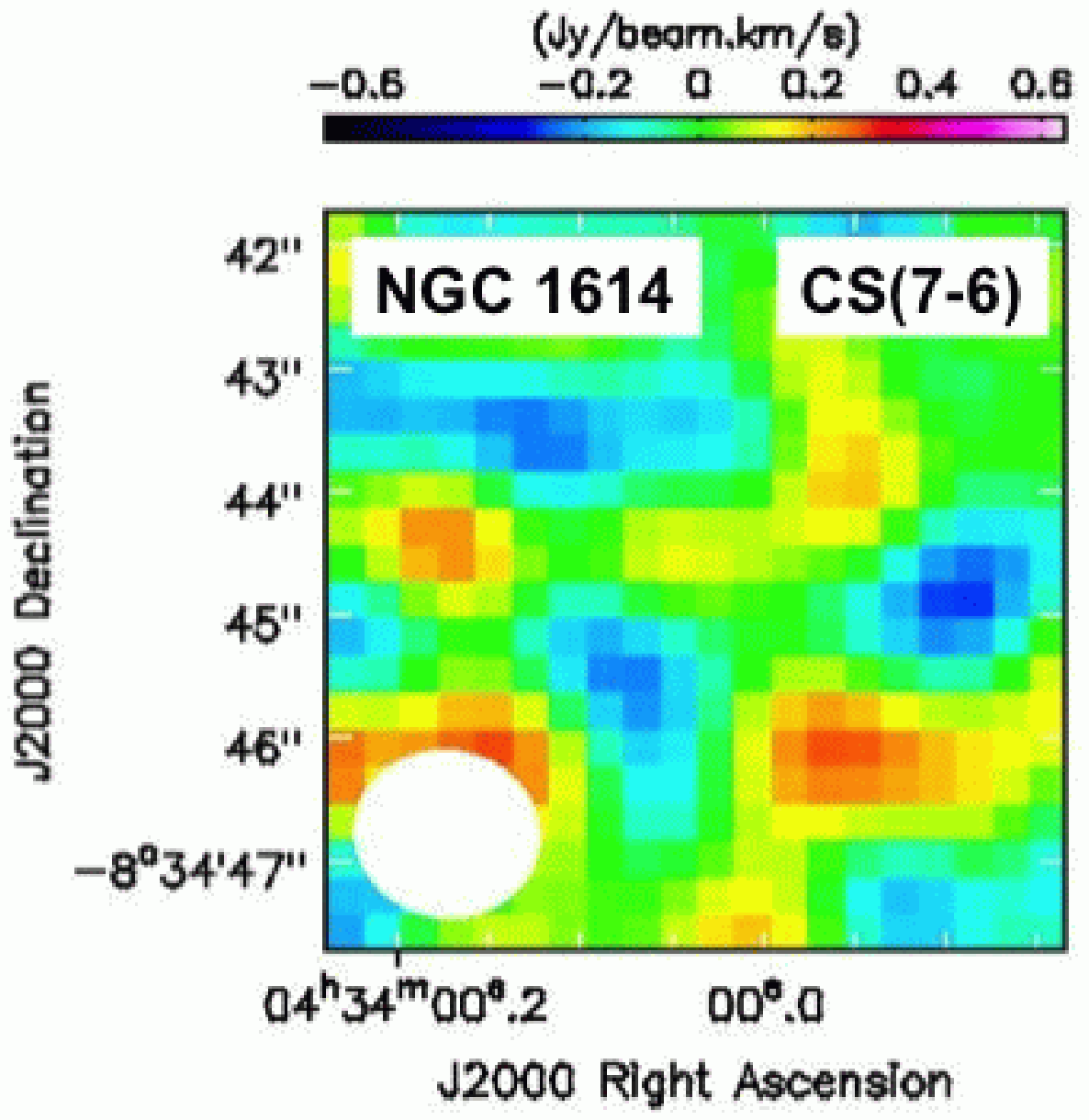} 
\includegraphics[angle=0,scale=.5]{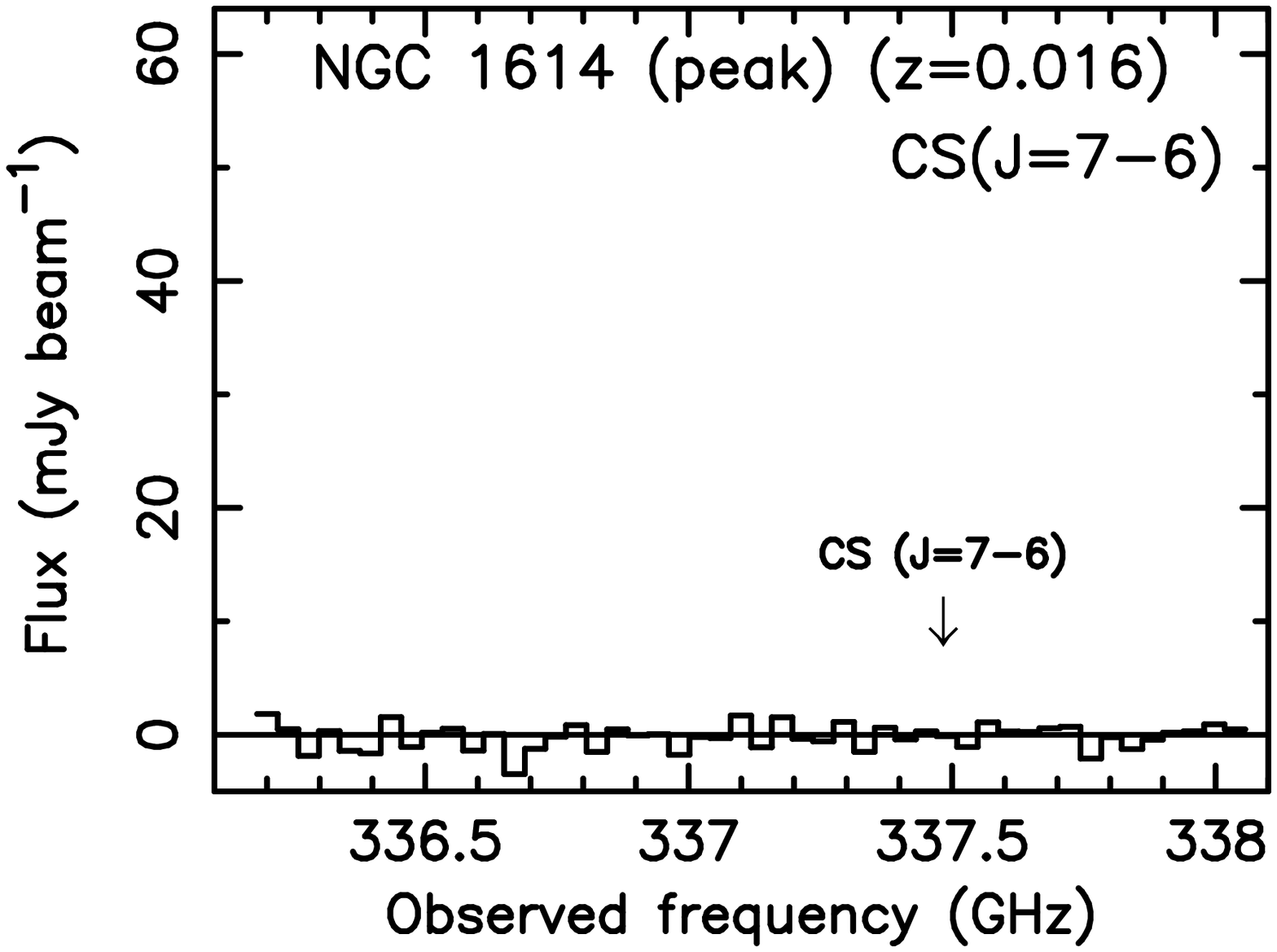}\\ 
\end{center}
\caption{
({\it Left}) :  
Integrated intensity (moment 0) maps of HCN J = 4--3, HCO$^{+}$ J = 4--3,
HNC J = 4--3, and CS J = 7--6 lines of NGC 1614. 
North is top, and east is to the left.
Signals in channels where line emission is recognizable are integrated
to produce the moment 0 maps.
Contours of the moment 0 maps are 
3$\sigma$, 5$\sigma$, 7$\sigma$, 9$\sigma$, and 11$\sigma$ for HCN,
5$\sigma$, 10$\sigma$, 15$\sigma$, 20$\sigma$, 25$\sigma$, and
30$\sigma$ for HCO$^{+}$, and 
3$\sigma$, 4$\sigma$, 5$\sigma$, and 6$\sigma$ for HNC.  
For CS J = 7--6, no emission feature with $\gtrsim$3$\sigma$ is seen.
The 1 $\sigma$ level is summarized in Table 4.
({\it Right}):  
The molecular line spectra within the beam size at the continuum peak
position are shown.
HCN J = 4--3, HCO$^{+}$ J = 4--3, and CS J = 7--6 line spectra are extracted
at the continuum ``a'' peak position (Table 3), and HNC J = 4--3 spectrum
is extracted at the continuum ``b'' peak position (Table 3).
The down arrows indicate the expected observed frequency of HCN J = 4--3, 
HCO$^{+}$ J = 4--3, HNC J = 4--3, and CS J = 7--6 lines at a redshift of
z = 0.016. 
In the HCO$^{+}$ J=4--3 line spectrum, the observed frequency of the
vibrationally excited HCN line (v$_{2}$ = 1, J = 4--3, l = 1f; 
$\nu_{\rm rest}$ = 356.256 GHz) \citep{sak10} is shown, 
but its detection is not clear.  
Assuming the same line profile as HCO$^{+}$ J=4--3, the flux of 
HCN J=4--3 at v$_{2}$=1 (l=1f) is estimated to be 
$<$0.45 [Jy beam$^{-1}$ km s$^{-1}$] ($<$3$\sigma$), 
which is $<$28\% of that of HCN J=4--3 at v=0.
}
\end{figure}

\begin{figure}
\includegraphics[angle=-0,scale=.45]{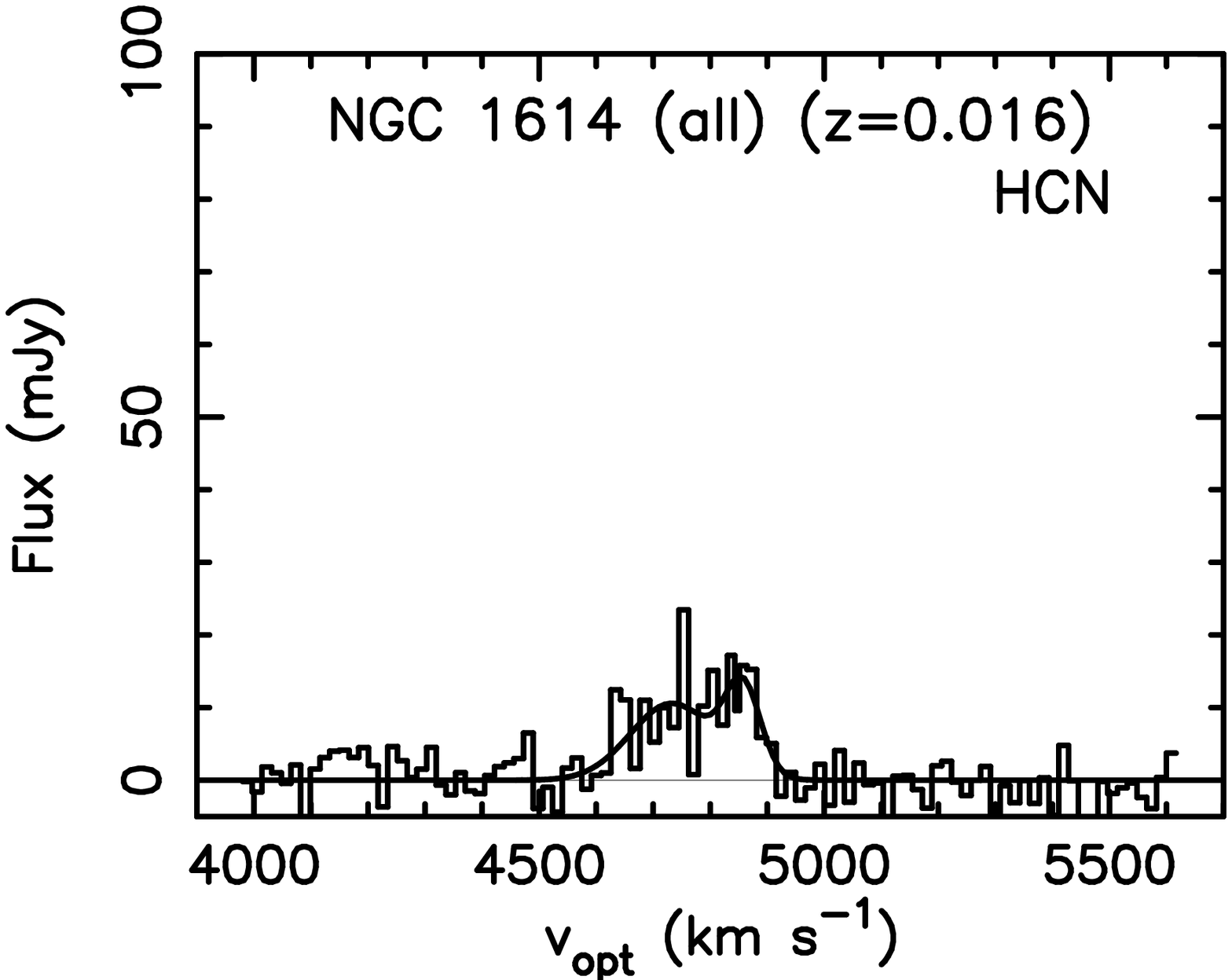} 
\includegraphics[angle=-0,scale=.45]{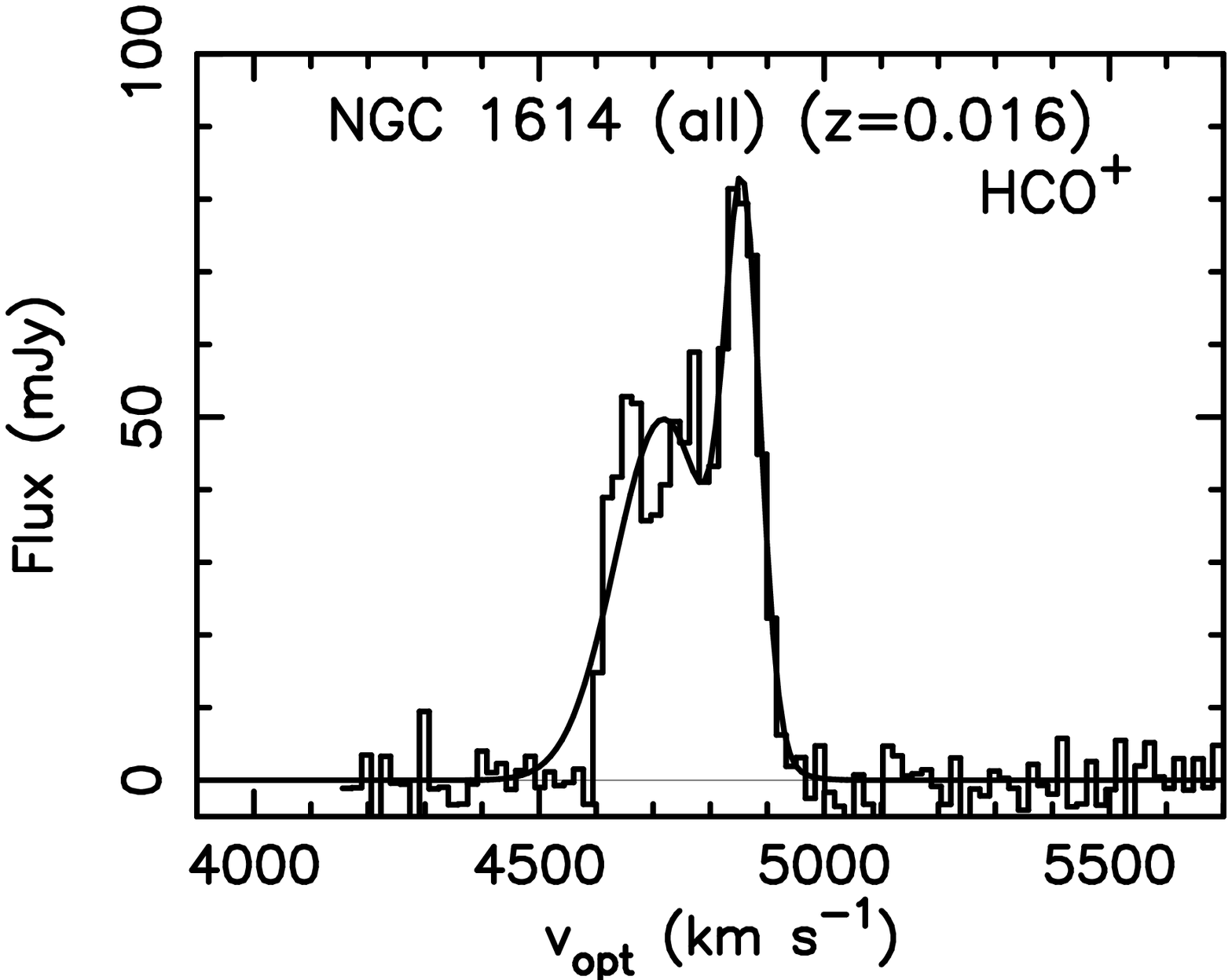} 
\includegraphics[angle=-0,scale=.45]{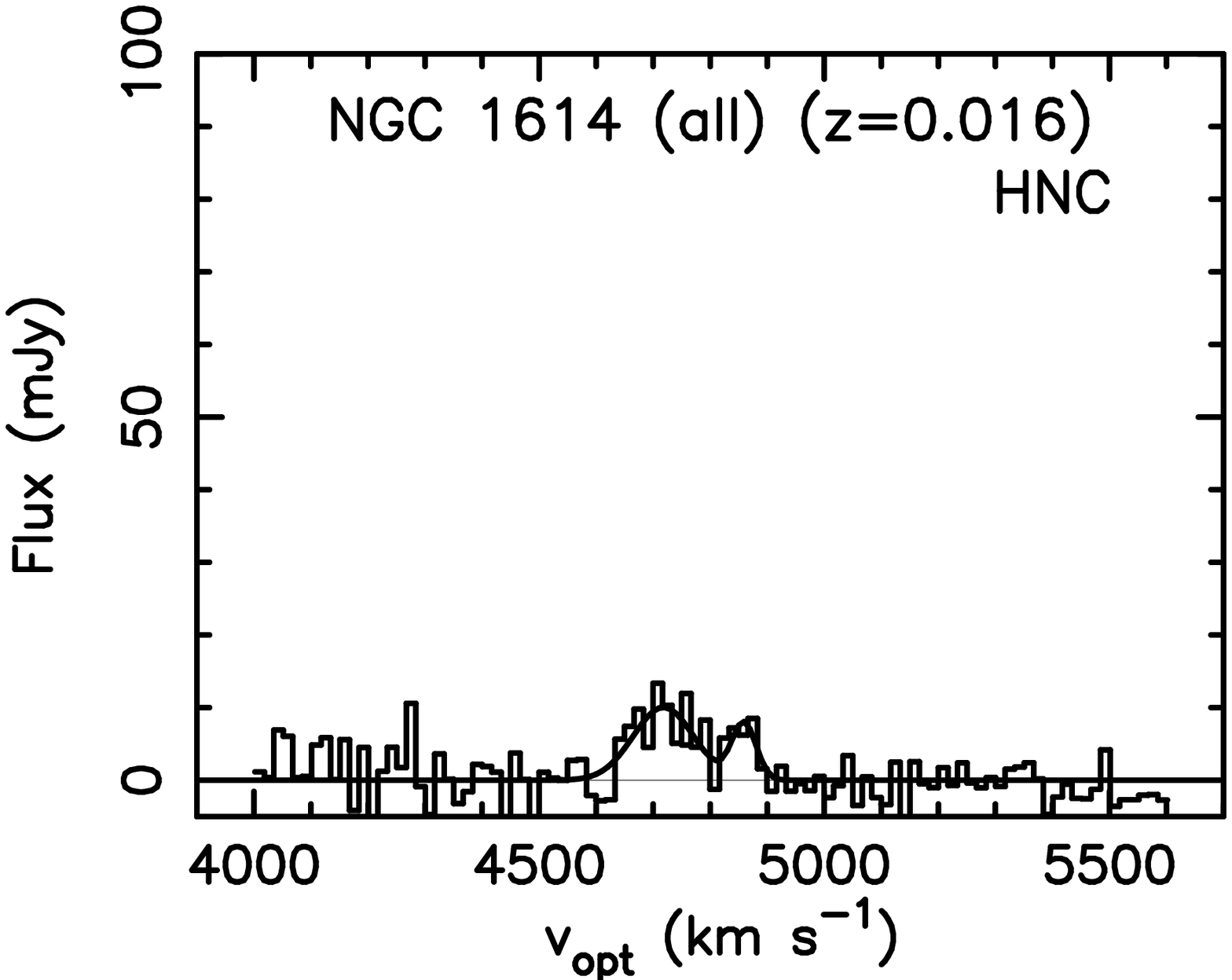} 
\includegraphics[angle=-0,scale=.45]{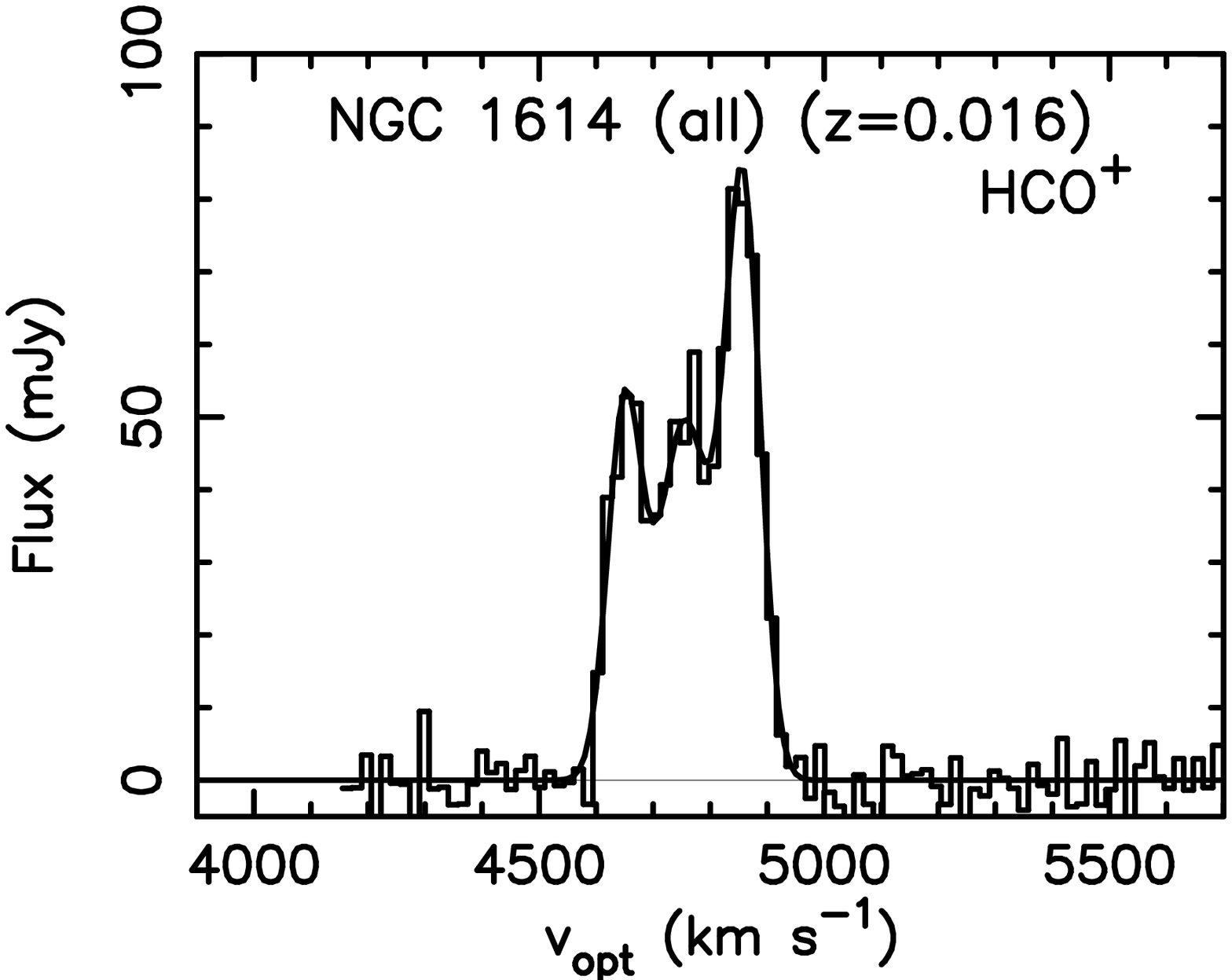} 
\caption{
HCN, HCO$^{+}$, and HNC J = 4--3 emission line spectra of NGC 1614
integrated over the region of significant signal detection
($\sim$3$''$ $\times$ 3$''$). 
Gaussian fits (Table 5) are overplotted as the solid curved line.
For the brightest HCO$^{+}$ J = 4--3 line, the triple Gaussian fit is also
overplotted (lower-right panel). 
}
\end{figure}

\begin{figure}
\includegraphics[angle=0,scale=.37]{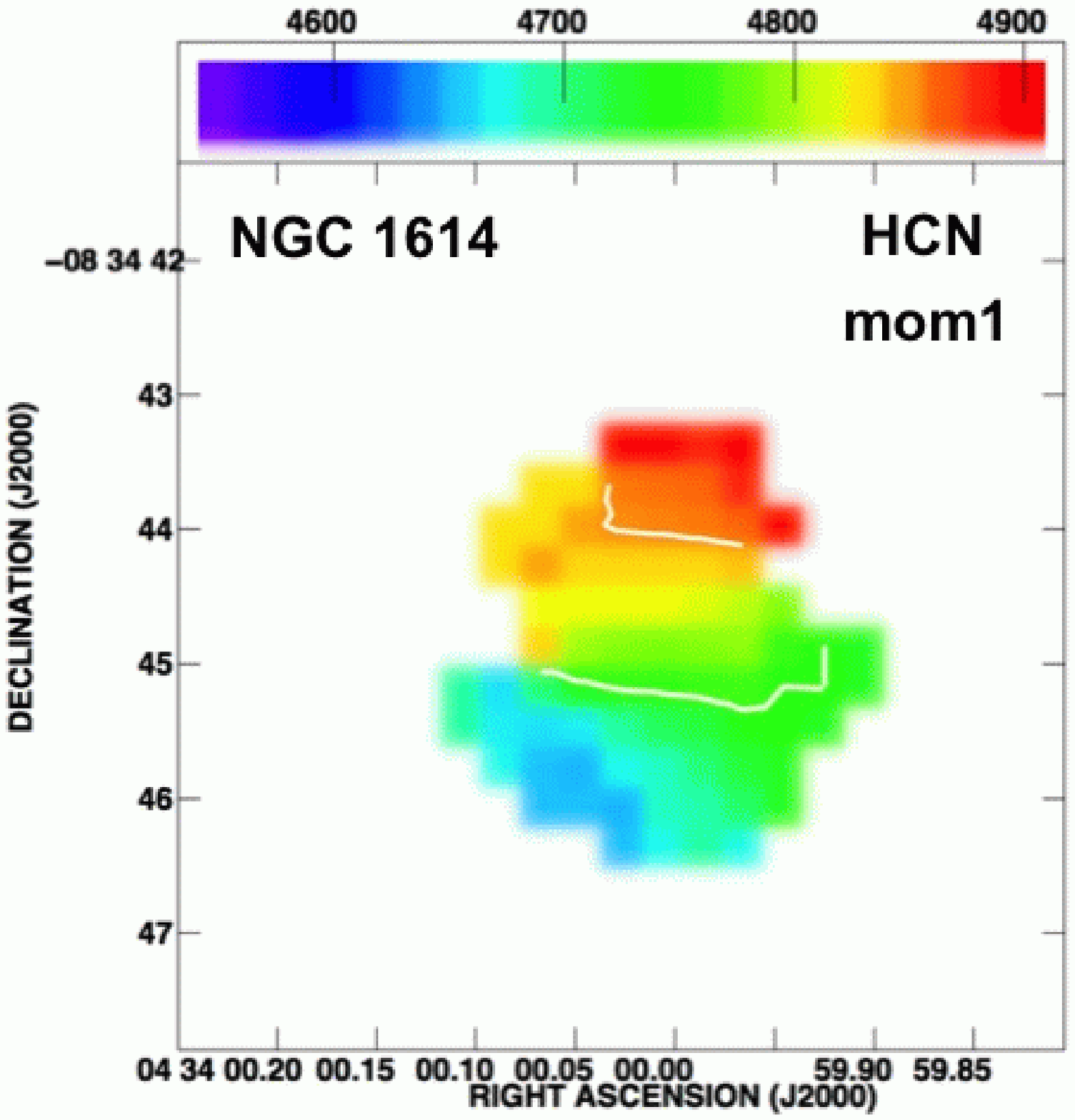} 
\includegraphics[angle=0,scale=.37]{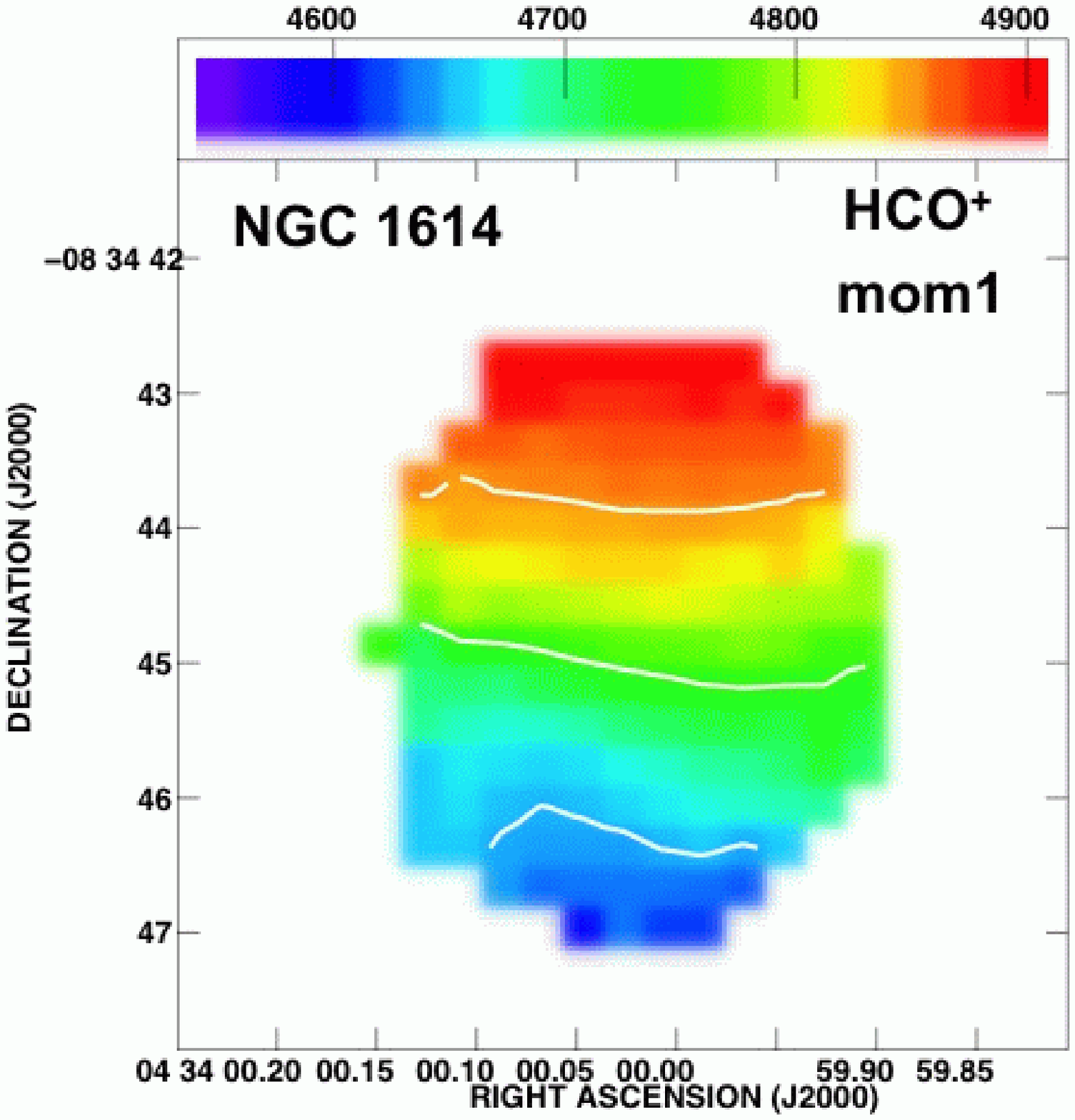} \\
\includegraphics[angle=0,scale=.37]{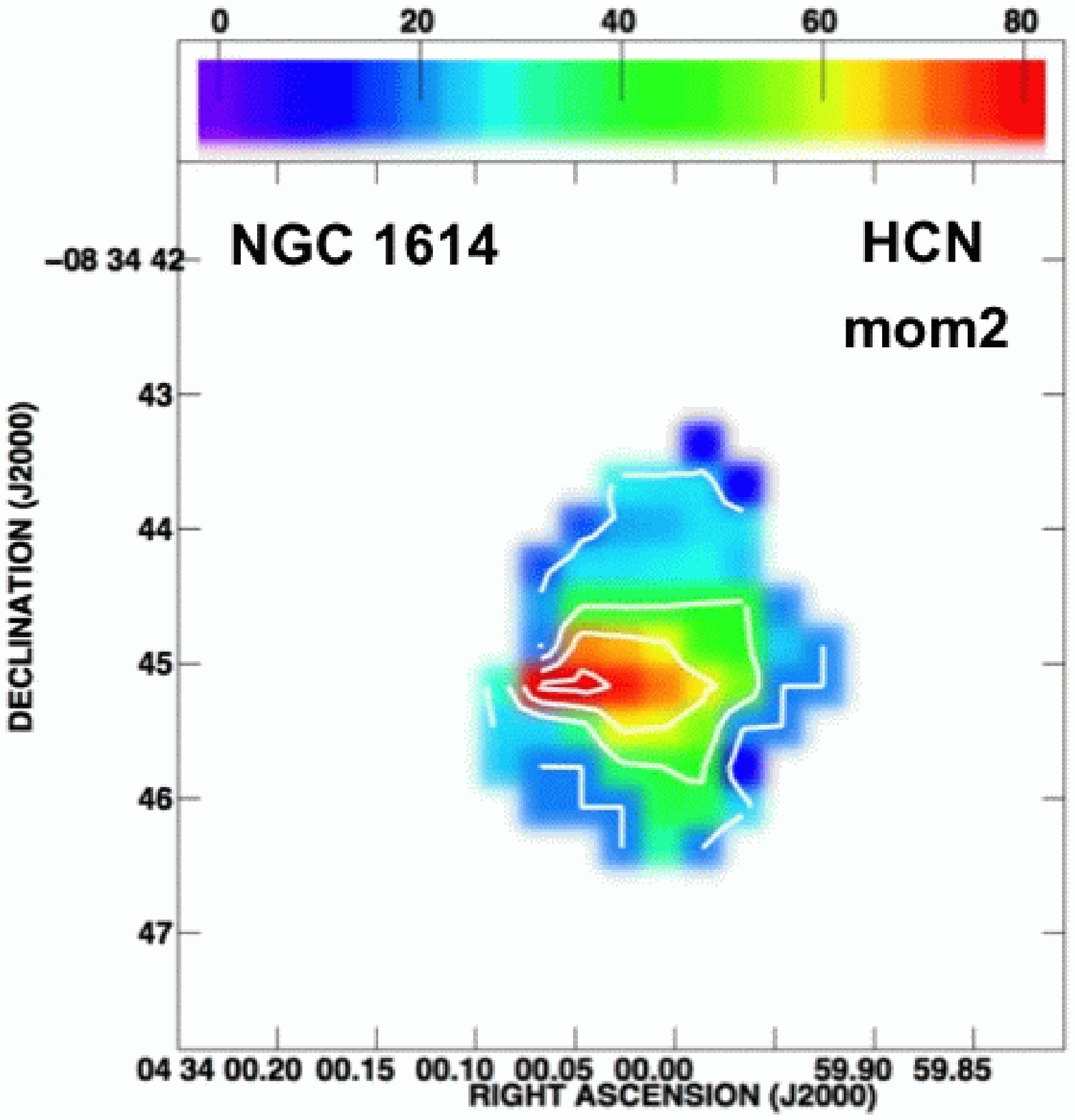} 
\includegraphics[angle=0,scale=.37]{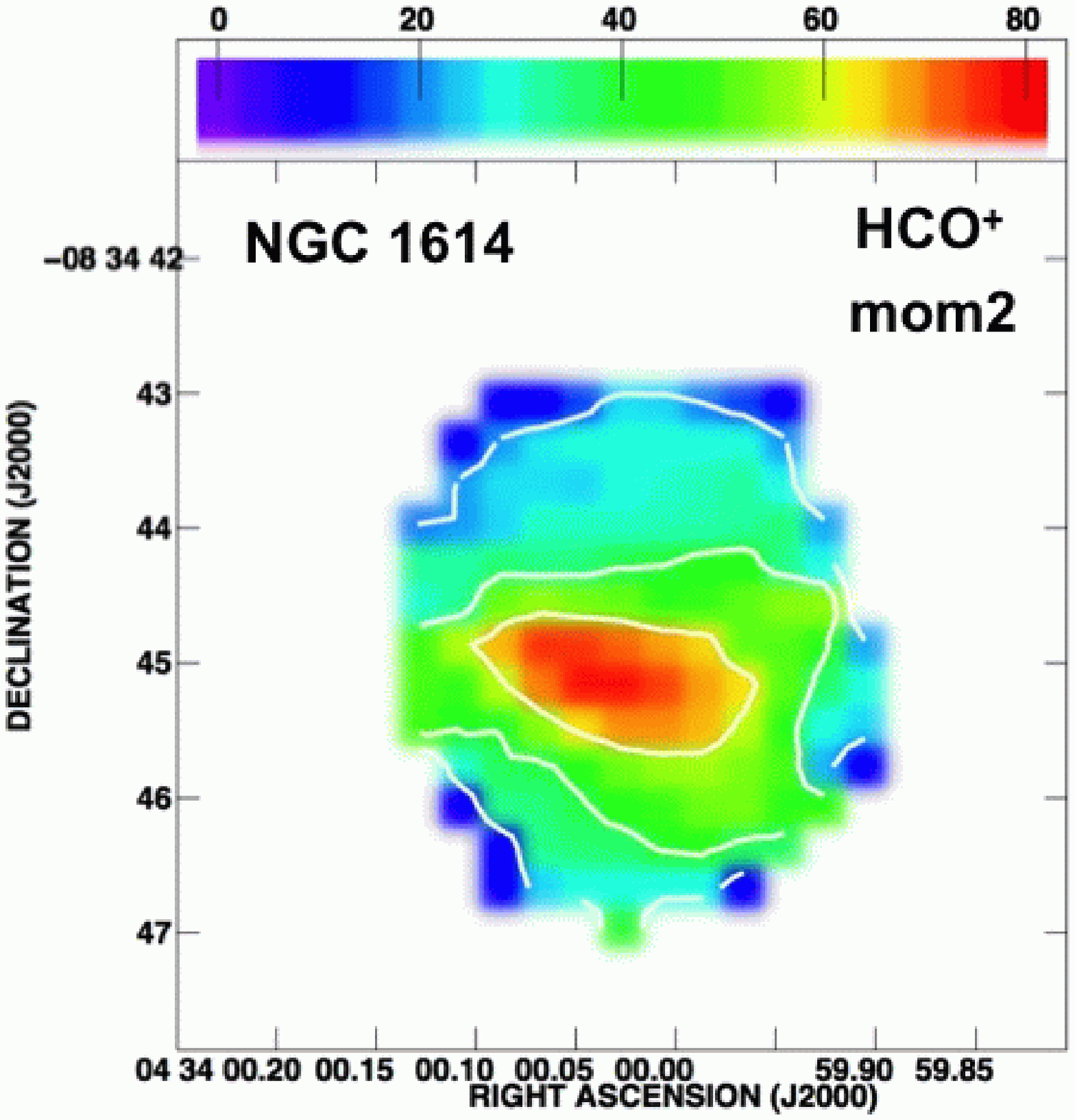} \\
\caption{
Intensity weighted mean velocity (moment 1) and intensity weighted
velocity dispersion (moment 2) maps of HCN and HCO$^{+}$ J = 4--3
emission lines from NGC 1614.  
The velocity is in v$_{\rm opt}$ 
$\equiv$ c ($\lambda$-$\lambda_{\rm 0}$)/$\lambda_{\rm 0}$. 
({\it Upper Left}): Moment 1 map of HCN J = 4--3 emission.
Contours are 4750 and 4850 [km s$^{-1}$]. 
({\it Upper Right}): Moment 1 map of HCO$^{+}$ J = 4--3 emission.
Contours are 4650, 4750, and 4850 [km s$^{-1}$].
({\it Lower Left}): Moment 2 map of HCN J = 4--3 emission.
Contours are 20, 40, 60, and 80 [km s$^{-1}$]. 
({\it Lower Right}): Moment 2 map of HCO$^{+}$ J = 4--3 emission.
Contours are 20, 40, and 60 [km s$^{-1}$]. 
}
\end{figure}

\begin{figure}
\includegraphics[angle=0,scale=.315]{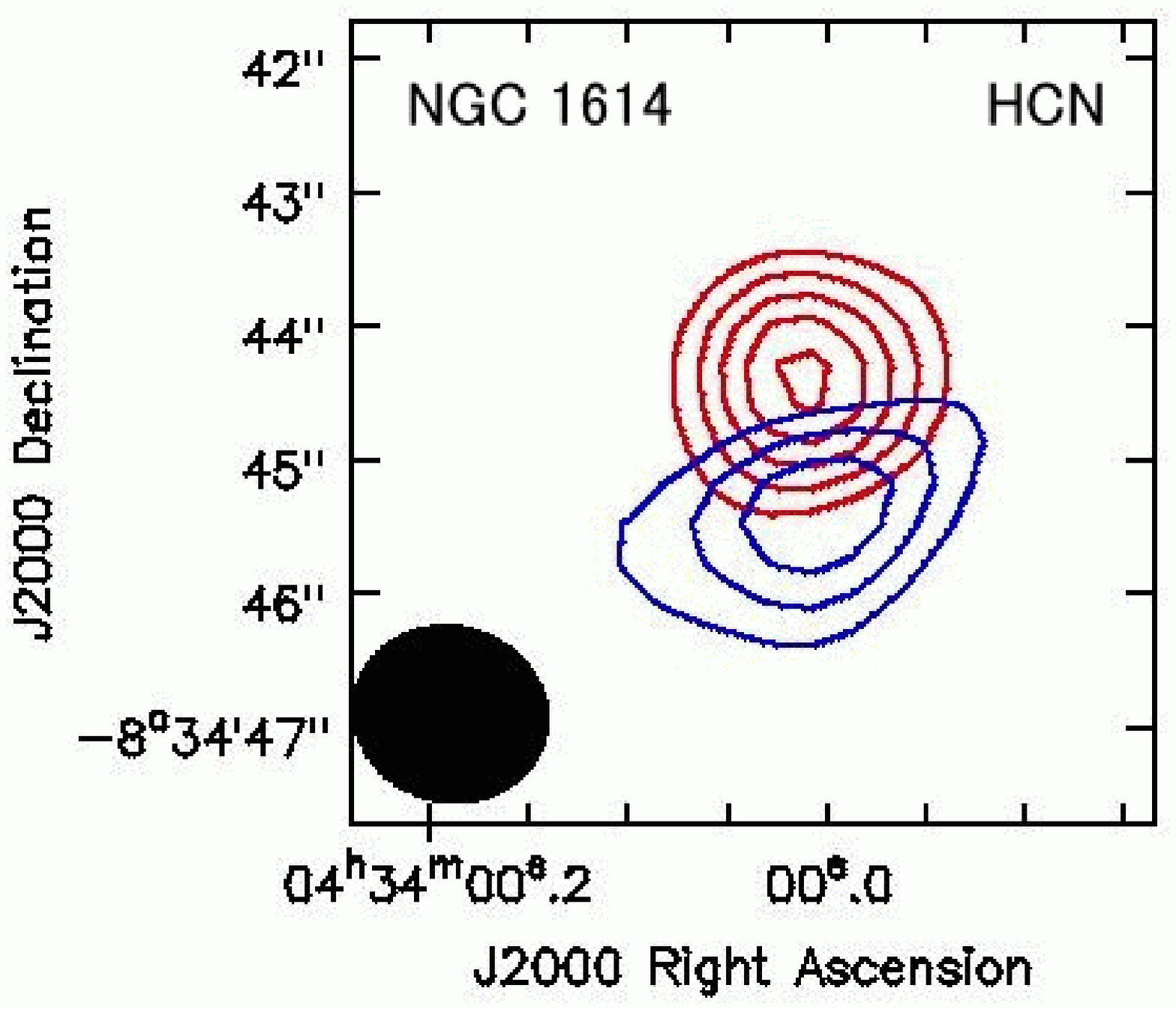} 
\includegraphics[angle=0,scale=.315]{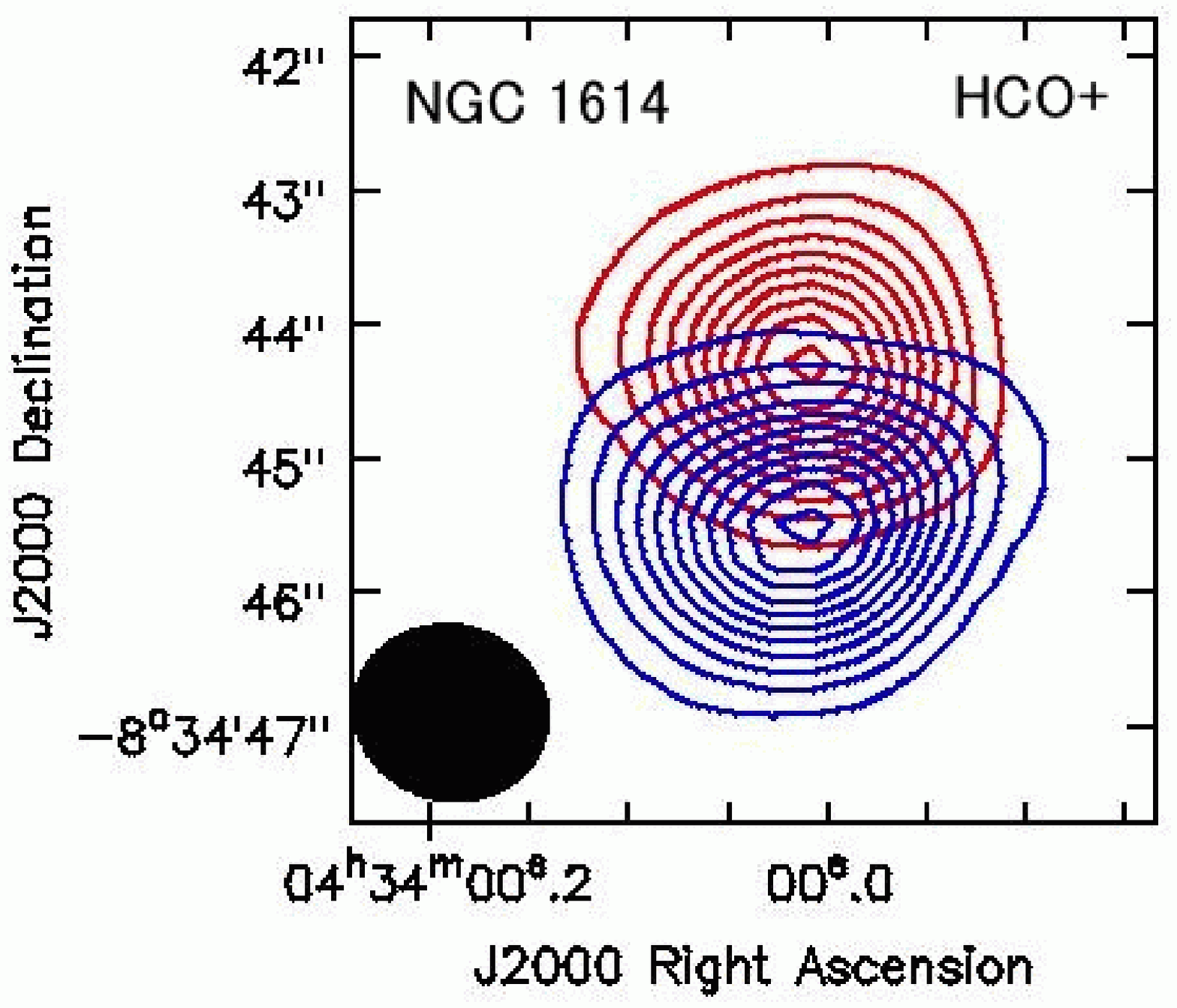} 
\includegraphics[angle=0,scale=.315]{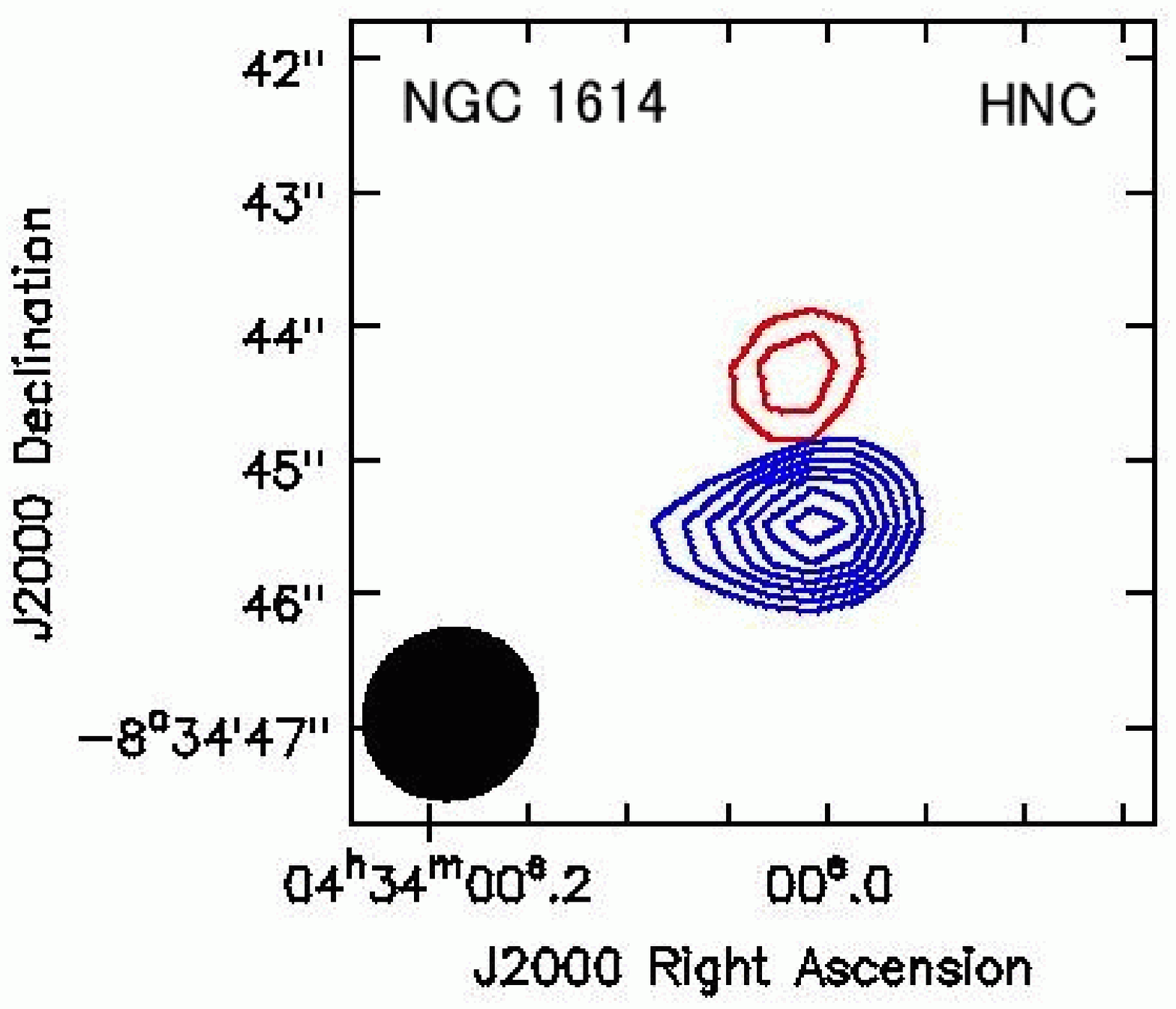} \\
\caption{
Contours of red and blue components of HCN, HCO$^{+}$, and HNC J = 4--3 
emission lines. 
By adopting a nuclear velocity for NGC 1614 (z = 0.016) of
v$_{\rm opt}$ = 4800 [km s$^{-1}$], emission with 
v$_{\rm opt}$ $>$ 4800 [km s$^{-1}$] and $<$ 4800 [km s$^{-1}$] 
is integrated for the red and blue components, respectively. 
The coordinates of the emission peaks of the red and blue components 
are (04 34 00.01, $-$08 34 44.3) and (04 34 00.01, $-$08 34 45.5) in
J2000, respectively, for all the HCN, HCO$^{+}$, and HNC J = 4--3 lines. 
For HCN, the contour starts at 0.4 [Jy km s$^{-1}$] and increases with 0.2
[Jy km s$^{-1}$].
For HCO$^{+}$, the contour starts at 0.5 [Jy km s$^{-1}$] and increases
with 0.5 [Jy km s$^{-1}$].
For HNC, the contour starts at 0.4 [Jy km s$^{-1}$] and increases with 0.1 
[Jy km s$^{-1}$].
}
\end{figure}

\begin{figure}
\includegraphics[angle=-0,scale=.45]{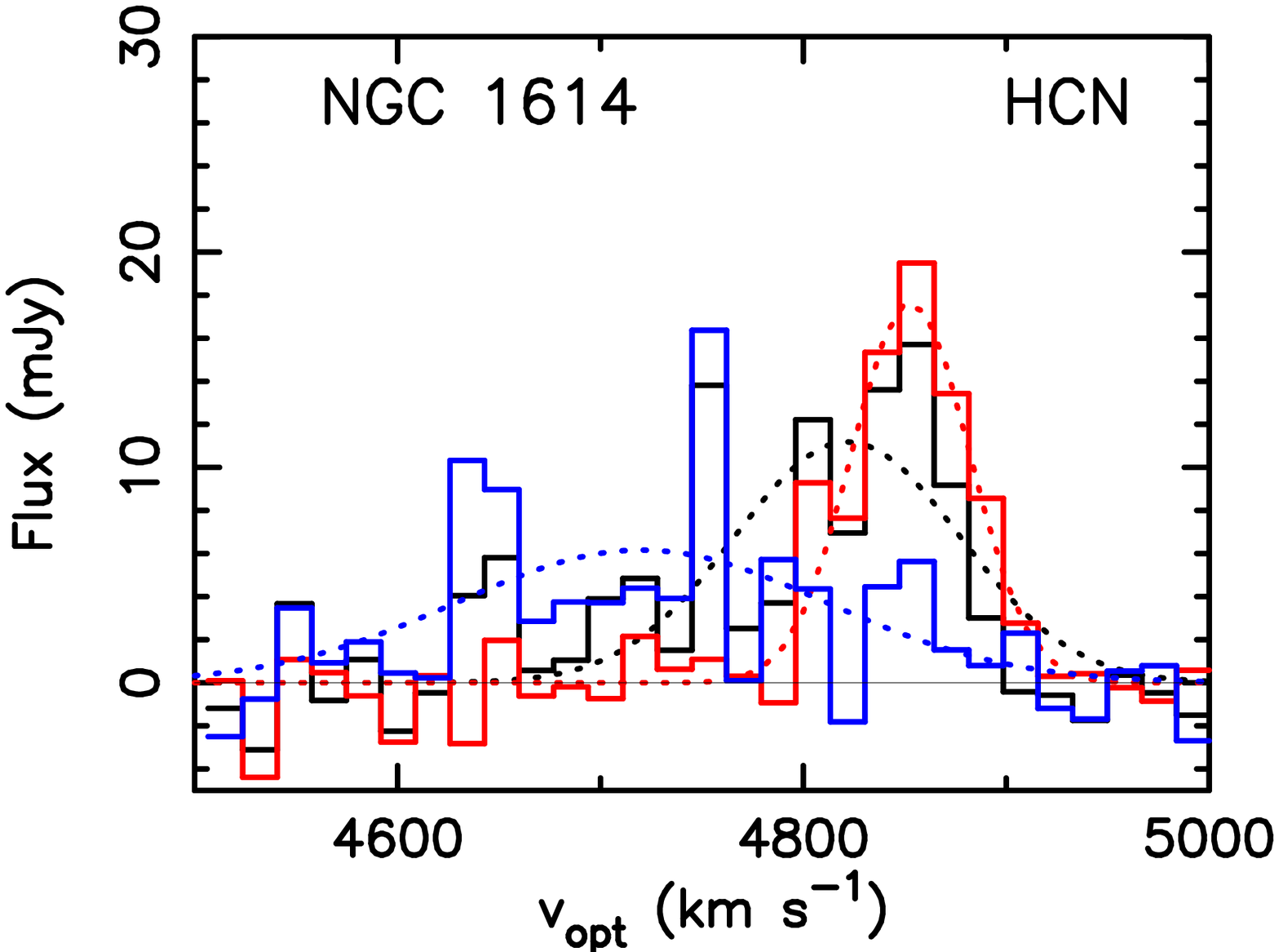} 
\includegraphics[angle=-0,scale=.45]{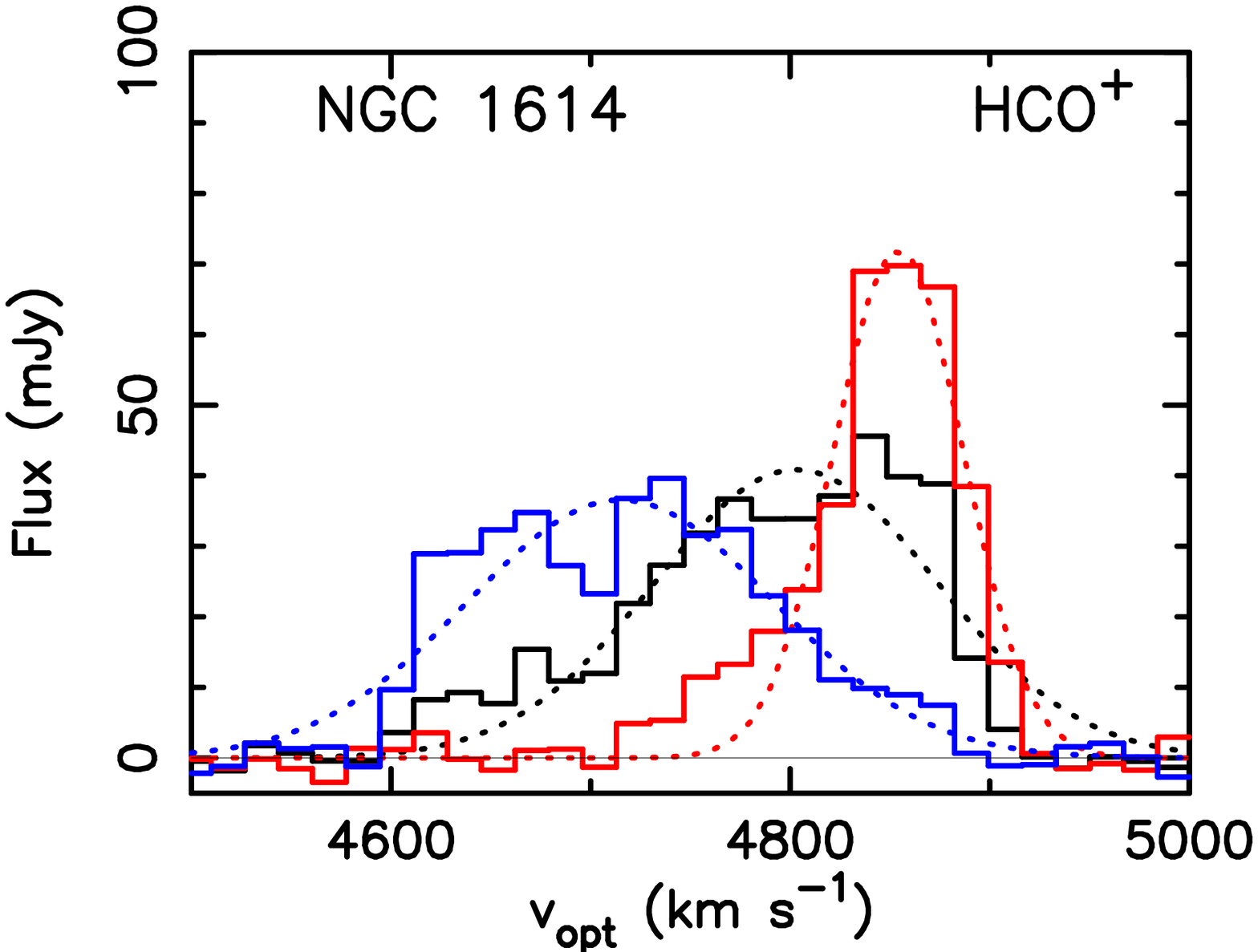} 
\includegraphics[angle=-0,scale=.45]{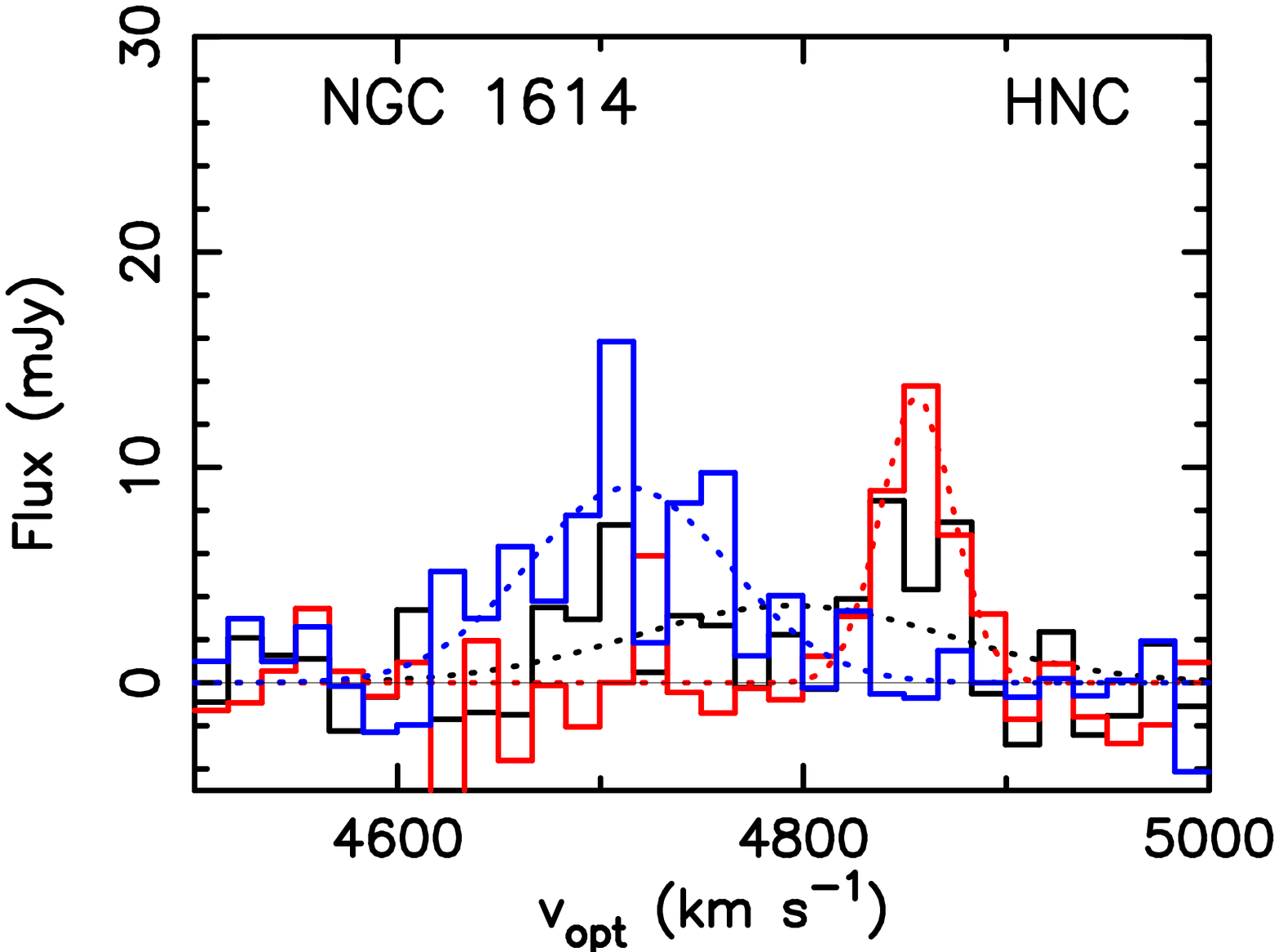} 
\caption{
Spectra of NGC 1614.
The black, red, and blue solid lines are spectra within the beam size at the nuclear position 
(04 34 00.01, $-$08 34 44.9) in J2000, the peak position of the 
red component (04 34 00.01, $-$08 34 44.3) in J2000, 
and the peak position of the blue component 
(04 34 00.01, $-$08 34 45.5) in J2000.
Gaussian fits are overplotted as dashed lines.
}
\end{figure}

\begin{figure}
\includegraphics[angle=-0,scale=.55]{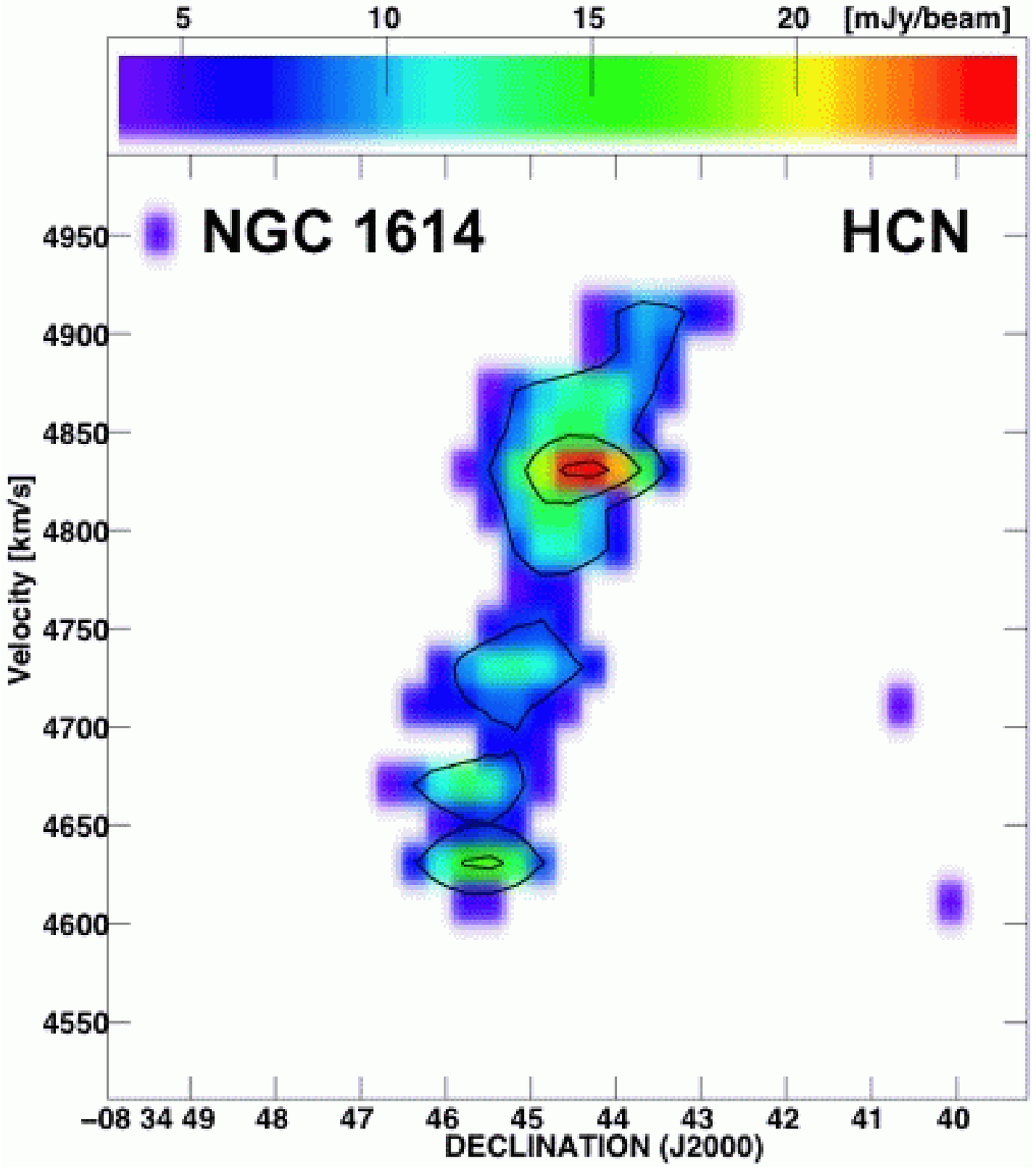} 
\includegraphics[angle=-0,scale=.55]{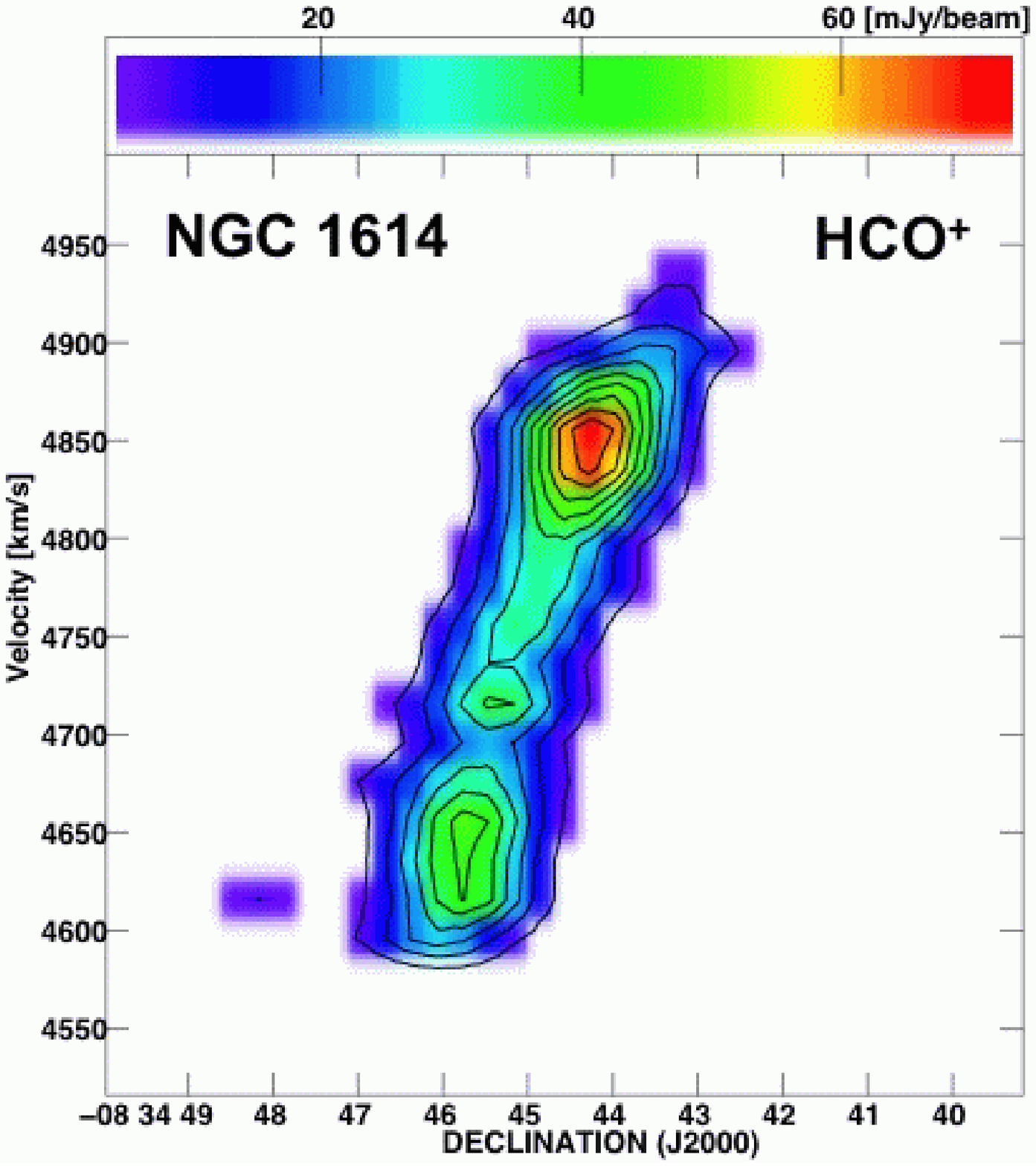}
\includegraphics[angle=-0,scale=.55]{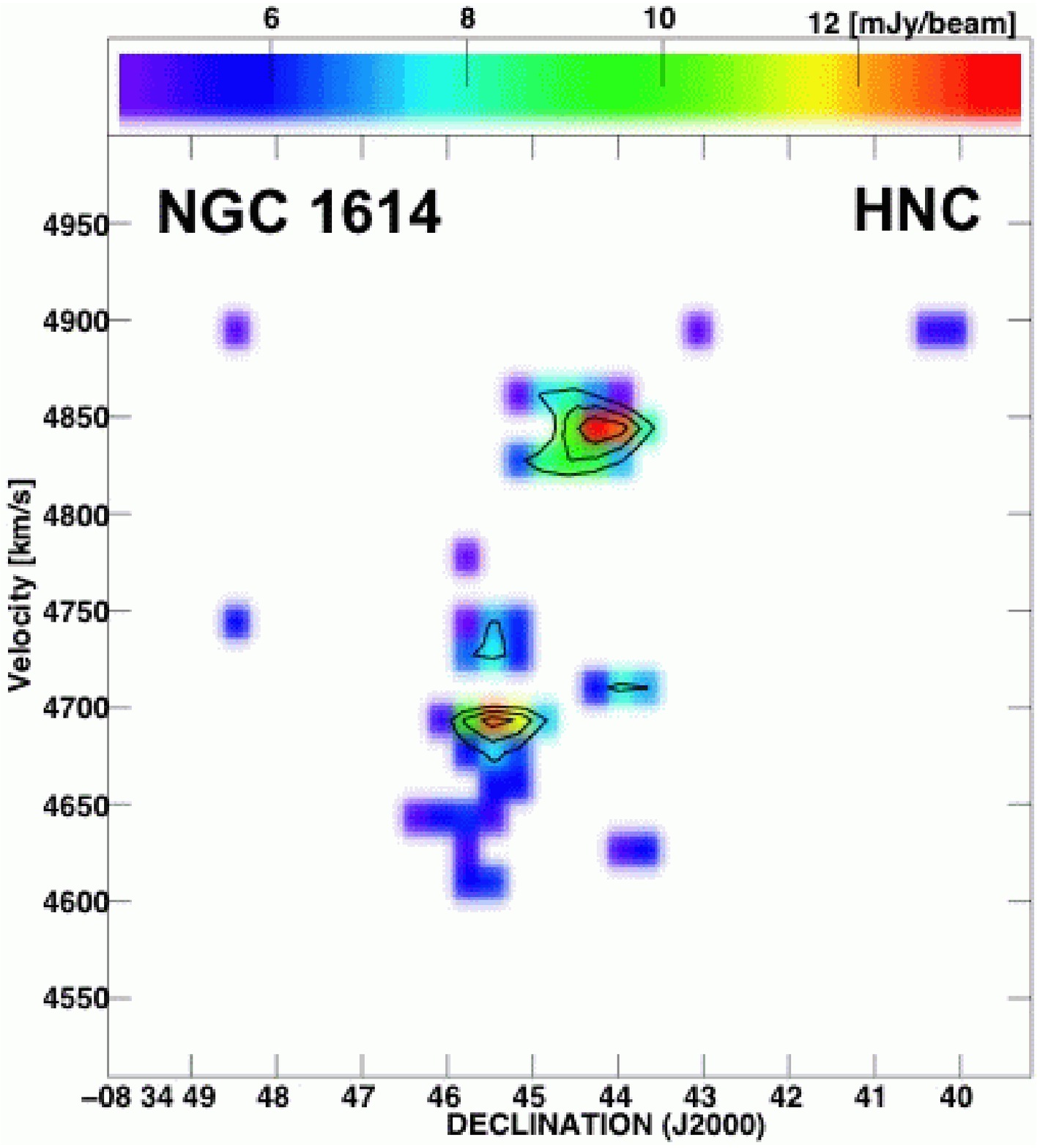} 
\caption{
A position-velocity diagram along the north-south direction 
through the nucleus (continuum "a" peak).
The abscissa is declination in J2000. 
North is to the right, and south is to the left.
The ordinate is optical LSR velocity.
The upper part of the y-axis is higher velocity, and the lower part is lower velocity.
For HCN, the contour starts at 7.6 [mJy beam$^{-1}$] and 
increases with 7.6 [mJy beam$^{-1}$].
For HCO$^{+}$, the contour starts at 7.2 [mJy beam$^{-1}$] and 
increases with 7.2 [mJy beam$^{-1}$].
For HNC, the contour starts at 4.6 [mJy beam$^{-1}$] and 
increases with 2.3 [mJy beam$^{-1}$].
}
\end{figure}

\begin{figure}
\begin{center}
\includegraphics[angle=-0,scale=0.9]{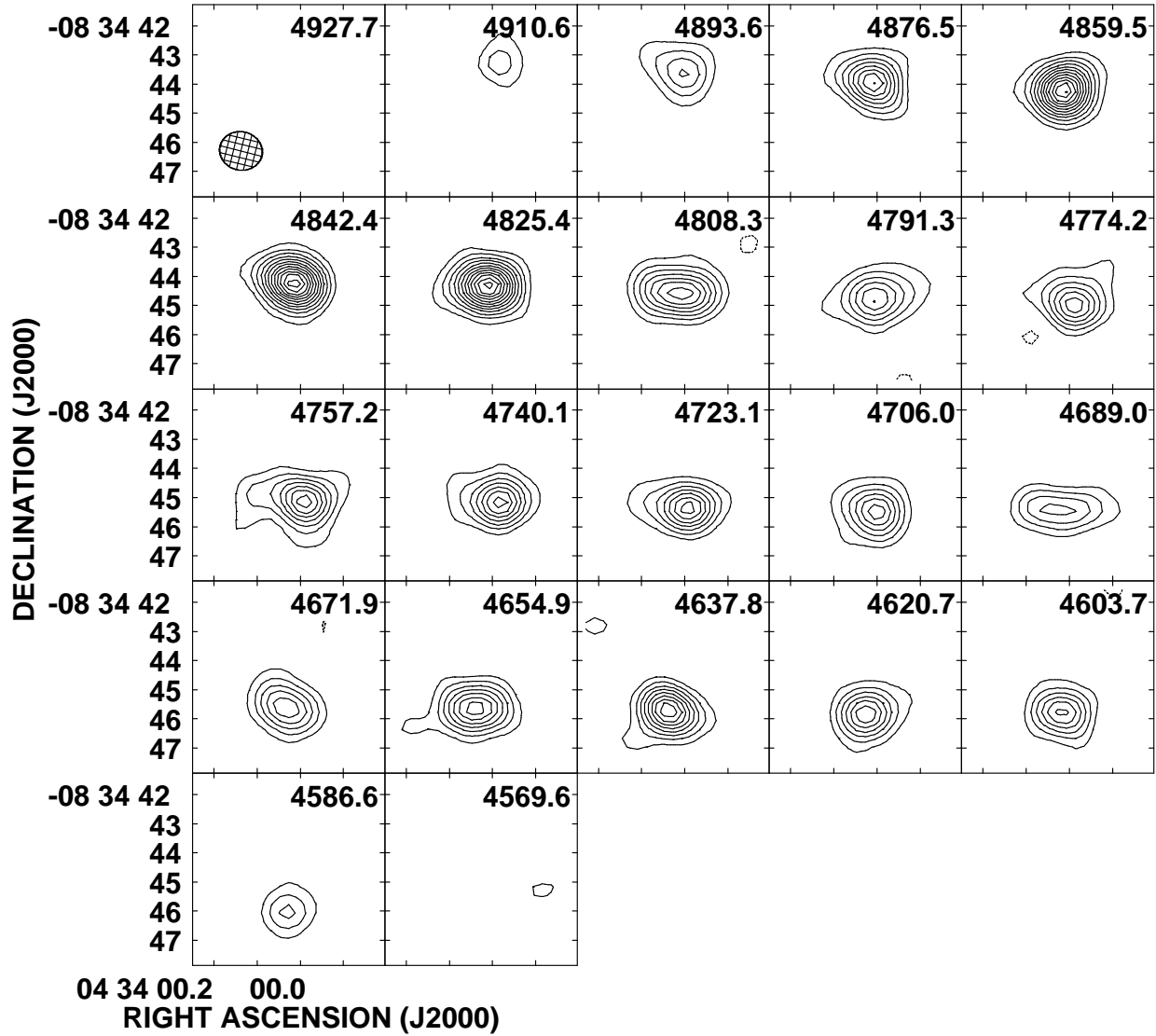} 
\end{center}
\caption{
A channel map of HCO$^{+}$ J = 4--3 emission in units of optical LSR
velocity. Contours start at 5.55 [mJy beam$^{-1}$] and increase
with 5.55 [mJy beam$^{-1}$]. 
The dashed contours are $-$5.55 [mJy beam$^{-1}$].
The rms noise level of each
channel is $\sim$1.85 [mJy beam$^{-1}$].
The number at the upper right part of each panel is velocity in 
[km s$^{-1}$], and the synthesized beam pattern is shown at the lower
left part of the top left panel. 
}
\end{figure}

\begin{figure}
\begin{center}
\includegraphics[angle=-0,scale=.45]{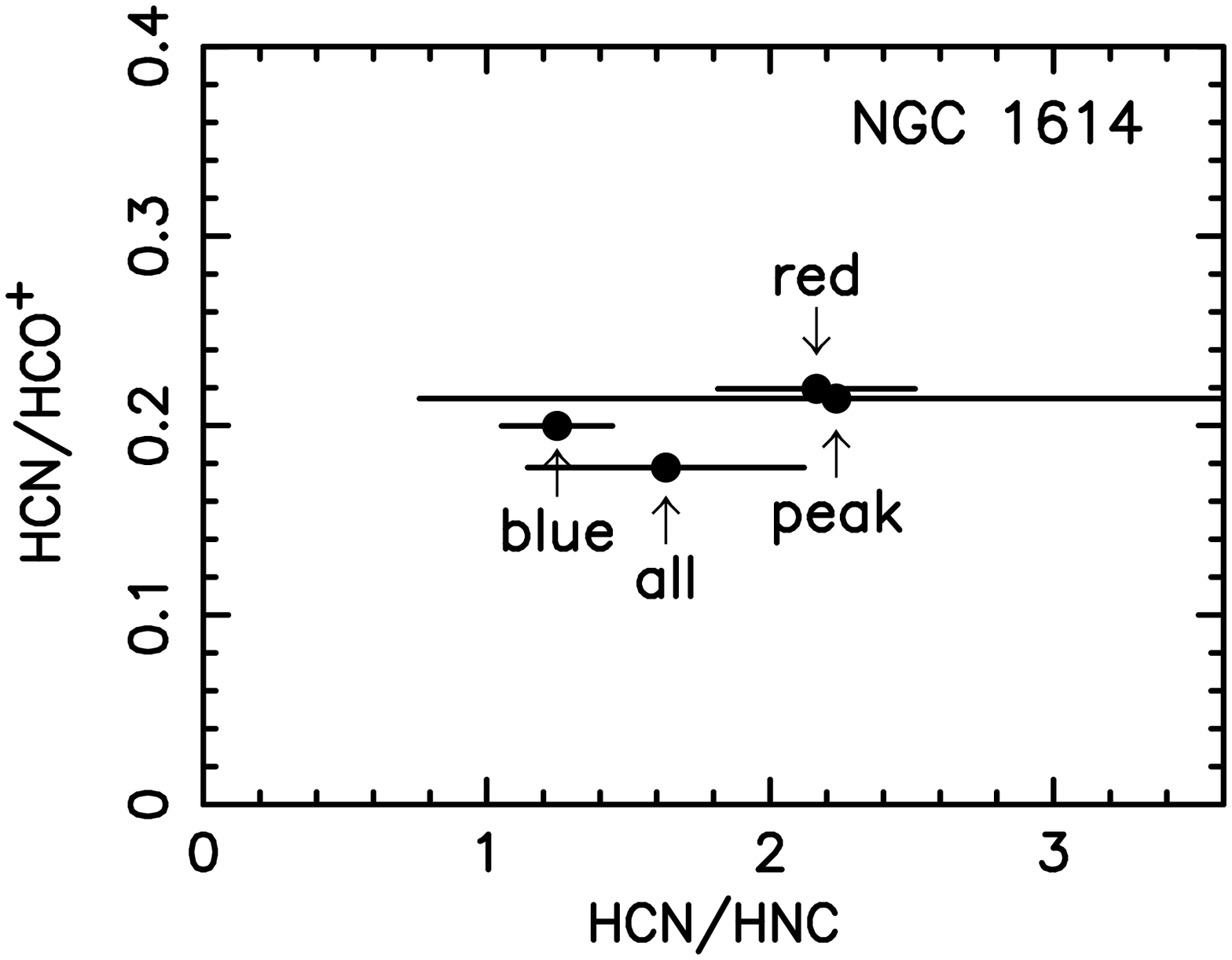} 
\end{center}
\caption{
The HCN-to-HNC flux ratio (abscissa) and HCN-to-HCO$^{+}$ flux ratio
(ordinate) at J = 4--3 transition. 
Data at ``all'', ``peak'', ``red'', and ``blue'' positions in Table 5
are used. 
}
\end{figure}


\begin{thebibliography}{}
\bibitem[Aalto et al.(1997)]{aal97}
         Aalto, S., Radford, S. J., E., Scoville, N. Z., \& Sargent,
         A. I. 1997, ApJ, 475, L107
\bibitem[Aalto et al.(2009)]{aal09}
         Aalto, S., Wilner, D., Spaans, M., Wiedner, M. C., Sakamoto,
         K., Black, J. H., \& Caldas, M. 2009, A\&A, 493, 481
\bibitem[Alonso-Herrero et al.(2001)]{alo01}
         Alonso-Herrero, A., Engelbracht, C. W., Rieke, M. J., Rieke,
         G. H., \& Quillen, A. C. 2001, ApJ, 546, 952
\bibitem[Bernard-Salas et al.(2009)]{ber09}
         Bernard-Salas, J., et al. 2009, ApJS, 184, 230 
\bibitem[Brandl et al.(2006)]{bra06}
         Brandl, B. R., et al. 2006, ApJ, 653, 1129   
\bibitem[Bryant \& Scoville(1999)]{bry99}
         Bryant, P., \& Scoville, N. Z. 1999, AJ, 117, 2632 
\bibitem[Casoli et al.(1999)]{cas99}
         Casoli, F., Willaime, M. -C., Viallefond, F., \& Gerin,
         M. 1999, A\&A, 346, 663
\bibitem[Costagliola et al.(2011)]{cos11}
         Costagliola, F., et al. 2011, A\&A, 528, 30
\bibitem[De Robertis \& Shaw (1988)]{der88}
         De Robertis, M. M., \& Shaw, R. A. 1988, ApJ, 329, 629
\bibitem[Diaz-Santos et al.(2008)]{dia08}
         Diaz-Santos, T., Alonso-Herrero, A., Colina, L., Packham, C., 
         Radomski, J. T., \& Telesco, C. M. 2008, ApJ, 685, 211   
\bibitem[Di Matteo et al.(2005)]{dim05}
         Di Matteo, T., Springel, V., \& Hernquist, L. 2005, Nature,
         433, 605
\bibitem[Downes \& Eckart(2007)]{dow07}
         Downes, D., \& Eckart, A. 2007, A\&A, 468, L57
\bibitem[Downes \& Solomon(1998)]{dow98}
         Downes, D., \& Solomon, P. M. 1998, ApJ, 507, 615 
\bibitem[Evans et al.(2002)]{eva02}
         Evans, A. S., Mazzarella, J. M., Surace, J. A., \& Sanders,
         D. B. 2002, ApJ, 580, 749 
\bibitem[Gao \& Solomon(2004)]{gao04} 
         Gao, Y., \& Solomon, P. M. 2004, ApJS, 152, 63
\bibitem[Haan et al.(2011)]{haa11}
         Haan, S., et al. 2011, AJ, 141, 100
\bibitem[Harada et al.(2010)]{har10}
         Harada, N., Herbst, E., \& Wakelam, V. 2010, ApJ, 721, 1570 
\bibitem[Hopkins et al.(2005)]{hop05}
         Hopkins, P. F., Hernquist, L., Cox, T. J., Di Matteo, T.,
         Martini, P., Robertson, B., \& Springel, V. 2005, ApJ, 630, 705
\bibitem[Hopkins et al.(2006)]{hop06}
         Hopkins, P. F., Hernquist, L., Cox, T. J., Di Matteo, T.,
         Robertson, B., \& Springel, V. 2006, ApJS, 163, 1
\bibitem[Imanishi \& Dudley(2000)]{imd00} 
         Imanishi, M., \& Dudley, C. C. 2000, ApJ, 545, 701 
\bibitem[Imanishi et al.(2006a)]{ima06a}
         Imanishi, M., Dudley, C. C., \& Maloney, P. R. 2006a, ApJ, 637, 
         114
\bibitem[Imanishi et al.(2011)]{ima11} 
         Imanishi, M., Imase, K., Oi, N., \& Ichikawa, K. 2011, AJ, 141,
         156
\bibitem[Imanishi et al.(2008)]{ima08}
         Imanishi, M., Nakagawa, T., Ohyama, Y., Shirahata, M., Wada,
         T., Onaka, T., \& Oi, N. 2008, PASJ, 60, S489
\bibitem[Imanishi \& Nakanishi(2006)]{in06} 
         Imanishi, M., \& Nakanishi, K. 2006, PASJ, 58, 813
\bibitem[Imanishi et al.(2004)]{ima04}
         Imanishi, M., Nakanishi, K., Kuno, N., \& Kohno, K. 2004, AJ,
         128, 2037 
\bibitem[Imanishi et al.(2006b)]{ima06b} 
         Imanishi, M., Nakanishi, K., \& Kohno, K. 2006b, AJ, 131, 2888
\bibitem[Imanishi et al.(2007)]{ima07} 
         Imanishi, M., Nakanishi, K., Tamura, Y., Oi, N., \& Kohno,
         K. 2007, AJ, 134, 2366
\bibitem[Imanishi et al.(2009)]{ima09} 
         Imanishi, M., Nakanishi, K., Tamura, Y., \& Peng, C. -H. 2009,
         AJ, 137, 3581 
\bibitem[Imanishi et al.(2010a)]{ima10a} 
         Imanishi, M., Nakanishi, K., Yamada, M., Tamura, Y., \& Kohno, K. 
         2010a, PASJ,   62, 201
\bibitem[Imanishi et al.(2010b)]{ima10b}
         Imanishi, M., Nakagawa, T., Shirahata, M., Ohyama, Y., \&
         Onaka, T. 2010b, ApJ, 721, 1233
\bibitem[Iono et al.(2007)]{ion07} 
         Iono, D., et al. 2007, ApJ, 659, 283
\bibitem[Iono et al.(2013)]{ion13} 
         Iono, D., et al. 2013, PASJ, in press (arXiv:1305.4535) 
\bibitem[Iono et al.(2004)]{ion04} 
         Iono, D., Ho, P. T. P., Yun, M. S., Matsushita, S., Peck,
         A. B., \& Sakamoto, K. 2004, ApJ, 616, L63 
\bibitem[Izumi et al.(2013)]{izu13}
        Izumi, T., et al. 2013, PASJ, in press (arXiv:1306:0507) 
\bibitem[Kewley et al.(2001)]{kew01}
         Kewley, L. J., Heisler, C. A., Dopita, M. A., \& Lumsden,
         S. 2001, ApJS, 132, 37 
\bibitem[Knudsen et al.(2007)]{knu07}
        Knudsen, K. K., Walter, F., Weiss, A., Bolatto, A., Riechers, D. A., \& Menten, K. 
        2007, ApJ, 666, 156
\bibitem[Kohno(2005)]{koh05}
         Kohno, K. 2005, in AIP Conf. Ser. 783, 
         The Evolution of Starbursts, ed. S. H\"uttemeister, E. Manthey,
         D. Bomans, \& K. Weis (New York: AIP), 203 (astro-ph/0508420)
\bibitem[Komatsu et al.(2009)]{kom09}
         Komatsu, E., et al. 2009, ApJS, 180, 330
\bibitem[Konig et al.(2013)]{kon13}
         Konig, S., Aalto, S., Muller, S., Beswick, R. J., \& Gallagher
         III, J. S. 2013, A\&A, 553, A72
\bibitem[Krips et al.(2008)]{kri08}
         Krips, M., Neri, R., Garcia-Burillo, S., Martin, S., Combes,
         F., Gracia-Carpio, J., \& Eckart, A. 2008, ApJ, 677, 262 
\bibitem[Lintott \& Viti(2006)]{lin06}
         Lintott, C., \& Viti, S. 2006, ApJ, 646, L37
\bibitem[Meijerink \& Spaans(2005)]{mei05}
         Meijerink, R., \& Spaans, M. 2005, A\&A, 436, 397 
\bibitem[Meijerink et al.(2007)]{mei07}
         Meijerink, R., Spaans, M., \& Israel, F. P. 2007, A\&A, 461, 793
\bibitem[Miles et al.(1996)]{mil96}
         Miles, J. W., Houck, J. R., Hayward, T. L., \& Ashby,
         M. L. N. 1996, ApJ, 465, 191  
\bibitem[Moorwood(1986)]{moo86}
         Moorwood, A. F. M. 1986, A\&A, 166, 4
\bibitem[Nakanishi et al.(2005)]{nak05}
         Nakanishi, K., Okumura, S. K., Kohno, K., Kawabe, R., \&
         Nakagawa, T. 2005, PASJ, 57, 575
\bibitem[Neff et al.(1990)]{nef90}
         Neff, S. G., Hutchings, J. B., Stanford, S. A., \& Unger,
         S. W. 1990, AJ, 99, 1088 
\bibitem[Olsson et al.(2010)]{ols10}
         Olsson, E., Aalto, S., Thomasson, M., \& Beswick, R. 2010,
         A\&A, 513, A11 
\bibitem[Perez-Beaupuits et al.(2007)]{per07}
         Perez-Beaupuits, J. P., Aalto, S., \& Gerebro, H. 2007, a\&A,
         476, 177 
\bibitem[Ridgway et al.(1994)]{rid94}
         Ridgway, S. E., Wynn-Williams, C. G., \& Becklin, E. E. 1994,
         ApJ, 428, 609
\bibitem[Rothberg \& Joseph(2004)]{rot04}
         Rothberg, B., \& Joseph, R. D. 2004, AJ, 128, 2098
\bibitem[Sakamoto et al.(2009)]{sak09}
         Sakamoto, K., et al. 2009, ApJ, 700, L104 
\bibitem[Sakamoto et al.(2013)]{sak13}
         Sakamoto, K., Aalto, S., Costagliola, F., Martin, S., Ohyama,
         Y., Wiedner, M. C., \& Wilner, D. J. 2013, ApJ, 764, 42
\bibitem[Sakamoto et al.(2010)]{sak10}
         Sakamoto, K., Aalto, S., Evans, A. S., Wiedner, M., \& Wilner,
         D. 2010, ApJ, 725, L228
\bibitem[Sanders et al.(2003)]{san03}
         Sanders, D. B., Mazzarella, J. M., Kim, D. -C., Surace, J. A., 
         \& Soifer, B. T. 2003, ApJ, 126, 1607
\bibitem[Sanders \& Mirabel(1996)]{sam96}
         Sanders, D. B., \& Mirabel, I. F. 1996, ARA\&A, 34, 749
\bibitem[Scoville et al.(1989)]{sco89}
         Scoville, N. Z., Sanders, D. B., Sargent, A. I., Soifer, B. T.,
         \& Tinney, C. G. 1989, ApJ, 345, L25
\bibitem[Sellgren(1981)]{sel81}
         Sellgren, K. 1981, ApJ, 245, 138
 \bibitem[Shier et al.(1994)]{shi94}
         Shier, L. M., Rieke, M. J., \& Rieke, G. H. 1994, ApJ 433, L9
\bibitem[Soifer et al.(2001)]{soi01}
         Soifer, B. T. et al. 2001, AJ, 122, 1213
\bibitem[Solomon et al.(1992)]{sol92}
         Solomon, P. M., Downes, D., \& Radford, S. J. E. 1992, ApJ,
         387, L55
\bibitem[Solomon \& Vanden Bout(2005)]{sol05}
         Solomon, P. M., \& Vanden Bout, P. A. 2005, ARA\&A, 43, 677
\bibitem[Springel et al.(2005)]{spr05}
         Springel, V., Di Matteo, T., \& Hernquist, L. 2005, MNRAS, 361,
         776 
\bibitem[Trung et al.(2001)]{tru01}
         Trung, D. V., Lo, K. Y., Kim, D. -C., Gao, Y., \& Gruendl,
         R. A. 2001, ApJ, 556, 141
\bibitem[Vaisanen et al.(2012)]{vai12}
         Vaisanen, P., Rajpaul, V., Zijlstra, A. A., Reunanen, J., \&
         Kotilainen, J. 2012, MNRAS, 420, 2209
\bibitem[Veilleux et al.(1995)]{vei95} 
         Veilleux, S., Kim, D. -C., Sanders, D. B., Mazzarella, J. M., \&
         Soifer, B. T. 1995, ApJS, 98, 171 
\bibitem[Wilson et al.(2008)]{wil08}
         Wilson, C. D., et al. 2008, ApJS, 178, 189 
\bibitem[Yuan et al.(2010)]{yua10}
         Yuan, T. -T., Kewley, L. J., \& Sanders, D. B. 2010, ApJ, 709,
         884 
\end{thebibliography}
\end{document}